\documentclass[prd,showpacs,preprint,eqsecnum,amsmath,amssymb,nofootinbib]{revtex4}
\usepackage{graphicx, tikz}
\usepackage{amssymb,amsmath, mathrsfs}
\usepackage{amssymb, graphics, setspace}
\usepackage{hyperref}
\usepackage[all]{xy}
\newcommand{\be}{\begin{equation}}
\newcommand{\ee}{\end{equation}}
\newcommand{\bea}{\begin{eqnarray}}
\newcommand{\eea}{\end{eqnarray}}
\newcommand{\bes}{\begin{subequations}}
\newcommand{\ees}{\end{subequations}}
\newcommand{\scri}{{\mathscr I}}
\newcommand{\la}{\langle}
\newcommand{\ra}{\rangle}
\newcommand{\w}{\omega}
\begin{document}
\title{Correlation patterns from massive phonons \\ in 1+1 dimensional acoustic black holes: A toy model}
\author{Richard~A.~Dudley}
\email{dudlra13@wfu.edu}
\author{Paul~R.~Anderson}
\email{anderson@wfu.edu}
\affiliation{Department of Physics, Wake Forest University, Winston-Salem, North Carolina 27109, USA}
\author{Roberto~Balbinot}
\email{balbinot@bo.infn.it}
\affiliation{Dipartimento di Fisica dell'Universit\`a di Bologna and INFN sezione di Bologna, Via Irnerio 46, 40126 Bologna, Italy\\
Centro Fermi - Museo Storico della Fisica e Centro Studi e Ricerche Enrico Fermi, Piazza del Viminale 1, 00184 Roma, Italy}
\author{Alessandro~Fabbri}
\email{afabbri@ific.uv.es}
\affiliation{Centro Fermi - Museo Storico della Fisica e Centro Studi e Ricerche Enrico Fermi, Piazza del Viminale 1, 00184 Roma, Italy \\ Dipartimento di Fisica dell'Universit\`a di Bologna and INFN sezione di Bologna, Via Irnerio 46, 40126 Bologna, Italy \\
Laboratoire de Physique Th\'eorique, CNRS UMR 8627, B\^at. 210, Universit\'e Paris-Sud 11, 91405 Orsay Cedex, France\\
Departamento de F\'isica Te\'orica and IFIC, Universidad de Valencia-CSIC, C. Dr. Moliner 50, 46100 Burjassot, Spain}
\begin{abstract}
Transverse excitations in analogue black holes induce a masslike term in the longitudinal mode equation. With a simple toy model we show that correlation functions display a rather rich structure characterized by groups of approximately parallel peaks.  For the most part the structure is completely different from that found in the
massless case.
\end{abstract}
\maketitle
\section{Introduction}

Recent years have witnessed a growing interest in the so-called analogue (gravity) models. These are condensed matter systems (Bose-Einstein condensates are the most studied) where one can mimic classical and quantum features present in a gravitational black hole (BH) or in cosmology \cite{Barcelo:2005fc}.

The major effort has been devoted to the study of the analogue of Hawking's quantum BH radiation \cite{hawking} that should appear in fluids when the flow undergoes a subsonic to supersonic transition \cite{Unruh}. This theoretical prediction has been confirmed for the first time by experiments performed by Steinhauer \cite{jeff1,jeff2}, who was able to catch in a Bose-Einstein condensate (BEC) the characteristic imprint in the correlation function of the Hawking pair creation \cite{paper1}, namely the presence of a peak.

Both a full quantum mechanical calculation~\cite{paper2} (see also \cite{Recati:2009ya, Macher:2009nz}) and one using a quantum field theory in curved space approach which includes backscattering
~\cite{paper2013}, predicted the existence of two further peaks beyond this primary one.

Most of the work for BEC analogue BHs assumed the flow to be one dimensional in which only longitudinal
modes are excited and the corresponding phonons are massless.
If one or more transverse modes are excited then a masslike term appears in the mode equation which is still a 1+1 dimensional equation.  Thus the phonons acquire a mass.  A preliminary investigation~\cite{mbh,Coutant:2012mf} showed hints of a much richer and more complex structure for the density density correlation function than occurs in the massless case.  In particular it was predicted that undulations should occur.

Undulations were first theoretically discovered for massless phonons in analogue white hole BECs in the supersonic region of the fluid \cite{whpaper}.  Undulations are zero frequency standing waves generated at the sonic horizon, the locus where the flow turns from subsonic to supersonic.
The Bogoliubov dispersion relation for a Bose-Einstein condensate (BEC) undergoing a one dimensional stationary flow along the negative $x$ direction is obtained by solving the Bogoliubov-de Gennes equation in the asymptotic regions where the flow speed and the sound speed are constant.\footnote{See for example Eqs. (2.5) and (2.6) of Ref.~\cite{paper2013}.  Note that the definition of the healing length $\xi$ there differs by a factor of two from that used here.}  It is given by
\be \label{dire} \omega -vk=\pm \sqrt{c^2k^2(1+\xi^2k^2)} \ee
where $\omega$ is the conserved frequency, $k$ the longitudinal wave vector, $c$ the speed of sound and
\be \xi = \frac{\hbar}{m_a c} \;, \label{healing} \ee
 the healing length with $m_a$ the mass of an atom in the BEC.  The $\pm$ signs refer to the positive or negative branch.

One sees that nontrivial zero frequency roots exist only if $v>c$.  They are
\be \label{cz} k_0=\pm \frac{1}{c_L\xi} \sqrt{v_L^2-c_L^2} \;, \ee
where $v_L$ and $c_L$ are the asymptotic ($x\to -\infty$) constant values of the velocity and the speed of sound in the supersonic region that we name $L$.
The existence of these roots relies on the dispersive character of Eq. (\ref{dire}).
Considering the group velocity $v_g=\frac{1}{\partial_\omega k}$ one can show that they are directed against the flow. This explains why the $k_0$ mode can be generated at the sonic horizon for white hole (WH) flows only.

However, when one allows for transverse excitations of the BEC, the dispersion relation changes to
\be \label{direm} \omega - vk=\pm \sqrt{c^2(k^2+k_\perp^2)[1+\xi^2(k^2+k^2_\perp)]}\ , \ee
where $k_\perp$ is the transverse wave vector.
One sees that for $k_\perp^2\xi^2\ll 1$ the quantity $c^2k_\perp^2$ acts as a mass squared term and in this regime we have two new zero frequency solutions of Eq. (\ref{direm});
for $v_L^2/c_L^2$  not too close to one they are
\be k_0^{m}=\pm \frac{c_L k_\perp}{\sqrt{v_L^2-c_L^2}}\ .\ee
The novelty is that these lie in the phonon (linear) branch of the dispersion relation and furthermore their group velocity is now directed along the flow.
This means that the presence of the masslike term induces undulations that can be generated at the sonic horizon in BH flows (not in WH) and that these are well described by the hydrodynamical approximation underlying the condensed matter-gravity analogy.

As shown in Ref.~\cite{whpaper}, in WH flows the undulation characterized by the wave vector $k_0$, Eq. (\ref{cz}), is responsible for the checkerboard pattern appearing when considering the BEC density-density correlation function when both points are located in the supersonic region of the WH flow.
One expects a similar behavior to emerge for BH flows inside the horizon, triggered by the presence of the transverse modes \cite{mbh}.

The motivation for this work was to investigate the details of the undulations using a simple model that should capture all the essential features of the underlying physics.
What we have found is that, while undulations exist, they do not dominate the behavior of the two-point correlation function.  Instead an equally complex structure of correlation
peaks is found.  It is shown that the most prominent peaks can be approximated using the stationary phase approximation
for the mode integral.

In Sec.~\ref{sec:model} the details of our model are given.  In Sec.~\ref{sec:unruh} explicit expressions for the mode functions and the two-point correlation function are given for the Unruh state.  Our results for the two-point correlation function are presented in Sec.~\ref{sec:results} along with a brief review of what happens in the massless case.
Section ~\ref{sec:conclusions} contains a brief summary and our conclusions.  One Appendix gives details of the relationship between our model and that of~\cite{paper1,paper2013}    and the other gives details of the normalization of some of the mode functions.

\section{The model}
\label{sec:model}

We shall consider a minimally coupled massive scalar field which satisfies the equation
\be (\Box -m^2)\hat\phi=0 \ , \label{kg-eq} \ee
where $\Box=\nabla_\mu\nabla^\mu$ is the covariant d'Alembertian.
The field propagates the 1+1 dimensional space-time described by the metric
\be \label{am} ds^2 = -c^2dT^2 + (dx + v_0 dT)^2 \;. \ee
We use this Painlev\'e-Gullstrand form for the metric because we wish to make a connection with a BEC analogue black hole.  In that case $\hat{\phi}$ would describe phase fluctuations in a BEC and the mass term would be related to the momentum $k_\perp$ of transverse excitations.  There are subtleties related to the connection which are discussed in Appendix~\ref{appendix-connection}.

In the BEC analogue the flow is assumed to be uniform in the negative $x$ direction with $\vec v=-v_0 \hat x$, $v_0 > 0$.  The speed of sound is assumed to vary uniformly along the same direction as the flow from an asymptotic value of $c_R$ for $x\to +\infty$ to an asymptotic value of $c_L$ for $x\to -\infty$, with
$c_R>v_0>c_L$.  The transition from subsonic to supersonic occurs at $x=0$ where $c=v_0$. The $L$ region ($x<0$) describes a sonic BH and the $R$ region ($x>0$) is outside the sonic horizon.
Typical functions used for the sound speed in the literature are $\tanh$ or $\tan^{-1}$. For the numerical computations in this paper we use the sound speed profile
\bea  c(x)&=&\sqrt{c_L^2 + \frac{1}{2}(c_R^2-c_L^2)\left[1+\frac{2}{\pi}\tan^{-1}\left(\frac{x+b}{\sigma_v}\right)\right]}  \;, \nonumber \\
      b &=& \sigma_v \tan \left[\frac{\pi}{c_R^2-c_L^2} \left( v_0^2 - \frac{1}{2} (c_R^2+c_L^2)\right) \right] \;, \label{c-used} \eea
which was also used in~\cite{paper2,paper2013}.  Here $\sigma_v$ is related to the width of the profile.

Throughout we use units such that $\hbar = c_R = 1$, although we will frequently include $c_R$ in expressions for clarity.  Lengths are
given in terms of the healing length evaluated in the limit $x \to \infty$ which is $\xi_1 = \frac{\hbar}{m_a c_R}$.  Masses are given in terms of the mass of a condensate atom, $m_a$ which is equivalent in
these units to $\xi^{-1}_1$.
For the plots of the two-point correlation function for massive phonons shown in Sec.~\ref{sec:results} we use the following values for the parameters:
\bea v_0 &=& \frac{3}{4}\;,  \qquad c_L = \frac{1}{2} \;, \qquad c_R = 1 \; \quad \text{and} \quad \sigma_v = 8 \xi_1  \; .  \label{c-parameters}  \eea
These are the same parameters that were used for the plots of the density density correlation function in~\cite{paper2013} in the massless case.

With standard manipulations one can rewrite the metric (\ref{am}) in Schwarzschild-like coordinates
\be \label{ams} ds^2=-(c^2-v_0^2)dt^2 + \frac{c^2}{c_2-v_0^2}dx^2 = - (c^2-v_0^2)(dt^2-dx^{*2})\ee
where
\be t=T-\int_{x_1}^x dy \frac{v_0}{c(y)^2-v_0^2} \;, \label{t-T} \ee
and
\be x^*=\int_{x_2}^x dy \frac{c(y)}{c(y)^2-v_0^2} \ . \label{xstar} \ee
Here $x_1$ and $x_2$ are arbitrary constants.  For the numerical calculations we use
\be x_1 = x_2 = \pm 10.5 \xi_1  \;.  \label{x1x2} \ee
Note that in $R$ the interval $(0,+\infty)$ in $x$ is mapped to $(-\infty,+\infty)$ in $x^*$, while in $L$ the interval $(-\infty,0)$ in $x$ is mapped to $(+\infty,-\infty)$ in $x^*$.

Recalling that the horizon is at $x = 0$, the surface gravity is given by
\be \kappa = \left(\frac{d c}{dx}\right)_{x=0}  \;. \label{kappa} \ee
For the parameters we are using
\be \kappa \xi_1 \approx 0.0185617 \;. \ee

In these coordinates the Klein Gordon equation~\eqref{kg-eq} becomes
\be \label{kgt} [ -\partial_t^2 +\partial_{x^*}^2 -m^2(c^2-v_0^2) ]\hat\phi =0\ . \ee
One should note that near the horizon the modes behave as effectively massless and that modes propagating in $R$ toward $x=+\infty$ are totally reflected back to the horizon when
$\omega < m\sqrt{c_R^2-v_0^2}$.

\begin{figure}[h]
\includegraphics[width=4in]{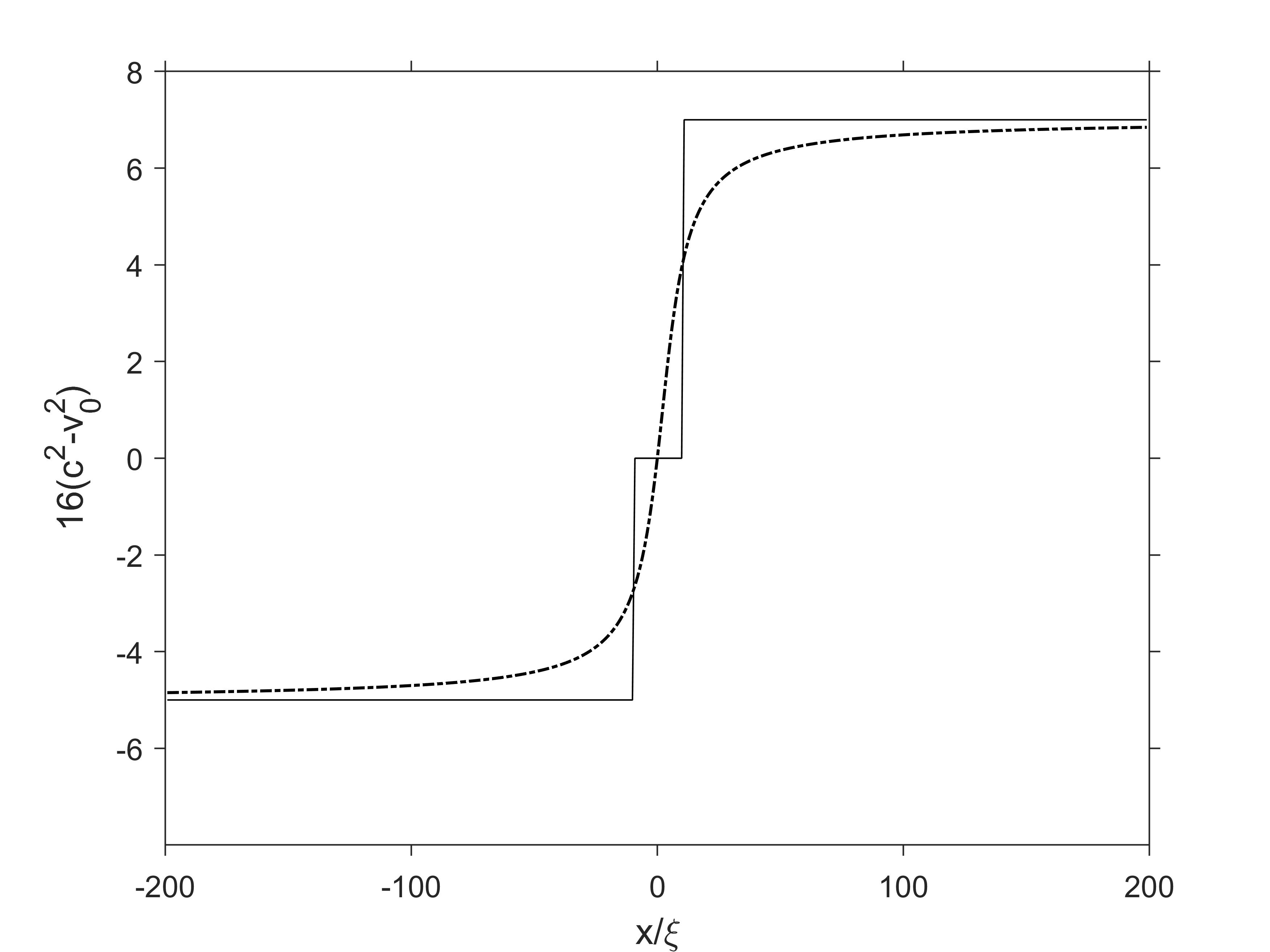}
\caption{\label{Fig:StepLikeApproximation}  Shown is a plot of the coefficient of $m^2$ in~\eqref{kgt} for the sound speed profile~\eqref{c-used} using the parameters~\eqref{c-parameters}(dashed)  and our approximation to that coefficient~\eqref{c2mv2-approx}(solid). }
\end{figure}

Given these considerations, we shall use a rather crude approximation for the masslike term in Eq.~(\ref{kgt}) which makes a simple analytical treatment possible while maintaining the essential physical behavior of the system. We approximate the mass term by a step function, both in $R$ and in $L$ so that
\bea  c^2 - v_0^2 &\to& (c_R^2-v_0^2) \theta(x^*-{}_Rx_0^*) \;, \qquad x > 0  \nonumber \\
&\to& (c_L^2-v_0^2)  \theta(x^*-{}_Lx_0^*) \;, \qquad x < 0  \;.  \label{c2mv2-approx} \eea
We choose ${}_Rx_0^*$ and ${}_Lx_0^*$ to be zero for simplicity.  For the numerical calculations the arbitrary constants in the definitions of $x^*$ in the $R$ and $L$ regions~\eqref{x1x2} were chosen so that~\eqref{c2mv2-approx} provides a reasonable approximation of $c^2 - v_0^2$ for the sound speed parameters used for our plots~\eqref{c-parameters};  see Fig.~\ref{Fig:StepLikeApproximation}.
Then Eq. (\ref{kgt}) simplifies to
\be \label{aet1}[-\partial_t^2 + \partial_{x^*}^2 -m_R^2 \theta(x^*-{}_Rx_0^*)]\hat \phi =0 \;, \ee
for $x>0$ (region $R$) with $m_R^2\equiv m^2(c_R^2-v_0^2)$, and
\be \label{aet2} [-\partial_t^2 + \partial_{x^*}^2 + m_L^2 \theta(x^*-{}_Lx_0^*)]\hat \phi =0 \;, \ee
for $x<0$ (supersonic region $L$) with $m_L^2\equiv m^2(v_0^2-c_L^2)$.
The classical solutions of Eqs.~(\ref{aet1}) and (\ref{aet2}) can therefore be expressed simply in terms of massless (where the step functions vanish) or massive plane waves.

\section{The Unruh state}
\label{sec:unruh}

\begin{figure}[h]
\begin{tikzpicture}
\node (I)    at ( 4,0)   {};
\node (II)   at (-4,0)   {};
\node (III)  at (0, 2.5) {};
\node (IV)   at (0,-2.5) {};
\node (V)   at (2.,-2.) {};
\node (VI)   at (6.25,-1.75) {};
\node (VII)    at ( 4.,2.5)   {$R$};
\node (VIII)   at (-2.,2.) {};

\node (IX)    at ( 0,4)   {$L$};

\path 
   (IX) +(90:4)  coordinate  (IXtop)
       +(-90:4) coordinate  (IXbot)
       +(180:4) coordinate (IXleft)
       +(0:4)   coordinate (IXright)
       ;
\draw (IXleft) --
          node[midway, below, sloped] {}
      (IXtop) --
          node[midway, above right]    {}
          node[midway, below, sloped] {}
      (IXright) --
          node[midway, below right]    {}
          node[midway, above, sloped] {}
      (IXbot) --
          node[midway, above, sloped] {}
      (IXleft) -- cycle;

\path  
  (II) +(90:4)  coordinate[label=90:]  (IItop)
       +(0:4)   coordinate                  (IIright)
       ;
\draw
      (IItop) --
          node[midway, below, sloped] {$H^-$}
      (IIright)-- cycle;

\path 
   (I) +(90:4)  coordinate[label=90:$i^+$]  (Itop)
       +(-90:4) coordinate[label=-90:$i^-$]  (Ibot)
       +(180:4) coordinate (Ileft)
       +(0:4)   coordinate[label=180:$i^0$] (Iright)
       ;
\draw (Ileft) --
          node[midway, below, sloped] {$\quad \quad H^+$}
      (Itop) --
          node[midway, above right]    {$\scri^+$}
          node[midway, below, sloped] {}
      (Iright) --
          node[midway, below right]    {$\scri^-$}
          node[midway, above, sloped] {}
      (Ibot) --
          node[midway, above, sloped] {}
      (Ileft) -- cycle;
\draw  (Ileft) -- (Itop) -- (Iright) -- (Ibot) -- (Ileft) -- cycle;

{\color{blue}\path[->]
 (V) +(45:0) coordinate[label=45:$\quad f_H^R$](IVtop)
	+(45:2) coordinate(IVbot)

	;
\draw[->] (IVtop) ->(IVbot);

\path[->]
 (VI) +(45:0) coordinate[label=90:](IVOut)
	+(135:1.5) coordinate(IVIn)

;

\path[->]
 (VIII) +(45:0) coordinate[label=45:$\quad f_H^L$](VIIItop)
	+(45:1.5) coordinate(VIIIbot)
;
\draw[->] (VIIItop) ->(VIIIbot);
}
\end{tikzpicture}
\caption{\label{Fig:PenroseNoScattering}Penrose diagram with modes which are nonzero on ${H_-}$ in the interior, ${L}$ and exterior, ${R}$.   }
\end{figure}
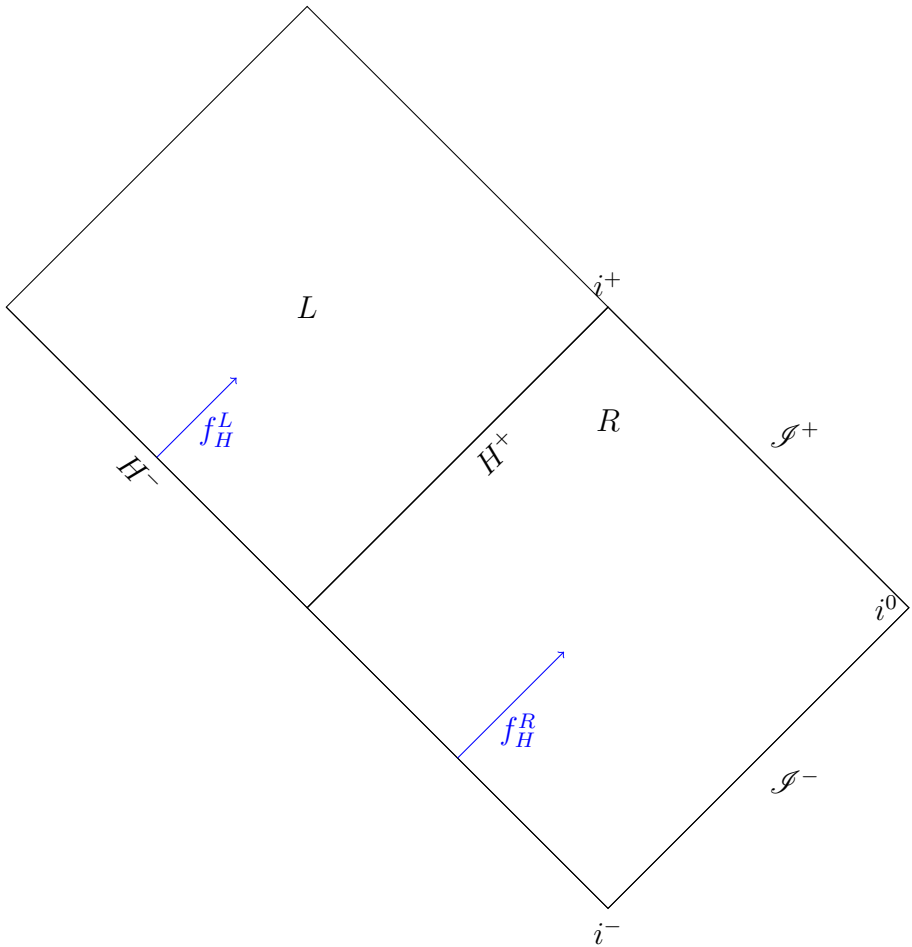

Hawking's BH evaporation can be understood as a pair production phenomenon triggered by the formation of a horizon in the gravitational collapse of a star leading to a BH.
Unruh \cite{unruh76} showed that the late time behavior of this dynamical process can be mimicked in a stationary BH metric by imposing appropriate boundary conditions
on the quantum state of the field on the past horizon of the Kruskal analytical extension of the BH metric.
In our case the corresponding Penrose diagram is given in Fig. \ref{Fig:PenroseNoScattering}.

For the Unruh state, one requires the modes, which we denote by $f_H^K(\omega_K)$, originating on the past horizon $H^-$ to be positive frequency with respect to the Kruskal $U$ coordinate defined as
\be \label{uk} U_K=\pm \frac{e^{-\kappa u}}{\kappa} \;. \ee
Here $-$ and $+$ refer to $R$ and $L$ respectively, and
\be u=t-x^* \ee
is an Eddington-Finkelstein (E.F.) retarded null coordinate.  Also needed are
the modes which propagate downstream from $x=+\infty$.  We denote them by $f_I(\omega)$ and they are positive frequency with respect to the time $t$.
Note that, given the supersonic character of region $L$, there are no modes coming from $x=-\infty$ and propagating upstream.

The field operator $\hat\phi$ is then expanded as
\be \label{exp} \hat \phi(x)=\int_0^{\infty} d\omega_K [\hat a_{\omega_K}f_H^K(\omega_K) + H.c.] + \int_{m_R}^{\infty} d\omega [\hat b_\omega f_I(\omega) + H.c.]\ . \ee
Here $\hat a_{\omega_K}$ and $\hat b_\omega$ are annihilation operators which together with their creation counterparts $\hat a_{\omega_K}^\dagger$ and
$\hat b_\omega^\dagger$ satisfy the usual boson commutation rules. Note the lower bound in the last integral: there are no incoming modes with $\omega<m_R$.

The modes are normalized as usual by the conserved scalar product
\be \label{sc} (f_1(\omega_1,x),f_2(\omega_2,x))= -i\int f_1(\omega_1,x)\overleftrightarrow{\partial_\mu} f_2^*(\omega_2 , x)[g_\Sigma (x)]^{\frac{1}{2}}d\Sigma^\mu  \ee
with $d\Sigma^\mu=n^\mu d\Sigma$, where $\Sigma$ is a Cauchy surface, $g_\Sigma$ the determinant of the induced metric and $n^\mu$ a future directed unit vector perpendicular to $\Sigma$.

Near the horizon ($x^*<{}_Rx_0^*$ in $R$ and $x^*<{}_Lx_0^*$ in $L$) the modes are massless. Therefore the correctly normalized modes emerging from the past horizon, according to the Unruh vacuum prescription, are
\be \label{mu} f_H^K(\omega_K)=\frac{1}{\sqrt{4\pi\omega_K}}e^{-i\omega_KU_K}\ . \ee
These modes are normalized on the past horizon, which is a Cauchy surface for these modes.

In $R$ the massless modes (\ref{mu}) propagate freely until they reach ${}_Rx_0^*$ where they encounter the step function barrier. They are partially (or totally) reflected back and partially transmitted becoming massive for $x^*>{}_Rx_0^*$.  
To find the form of the reflected and (eventually) transmitted part one has to impose junction conditions for the modes at the discontinuity point $x^*={}_Rx_0^*$.
As usual, one requires continuity of the modes and their first derivatives. Since the boundary is expressed in the $x^*$ coordinate, it is analytically complicated to rewrite the
$f_H^K(\omega_K)$ modes of Eq.~(\ref{mu}) in $(t,x^*)$ coordinates. It is much simpler to use a Bogoliubov transformation to expand these Unruh modes in terms of the usual
Eddington-Finkelstein retarded modes defined in $R$ and $L$ (see Fig.~\ref{Fig:PenroseNoScattering}).  On the past horizon they are
\be \label{rm} f_H^R=\frac{e^{-i\omega u}}{\sqrt{4\pi\omega}} \ , \ \ f_H^L=\frac{e^{i\omega u}}{\sqrt{4\pi\omega}} \;. \ee
Both of them are correctly normalized on the past horizon and positive frequency, $f^R_H$ with respect to the time $t$ and $f^L_H$ with respect to the
interior time $x^*$.
Therefore
\be \label{expu} f_H^K(\omega_K)= \int_0^\infty d\omega\left[\alpha_{\omega_K\omega}^Lf^L_H(\omega)+\beta_{\omega_K\omega}^Lf^{L*}_H(\omega) \right]+
\int_0^\infty d\omega\left[\alpha_{\omega_K\omega}^Rf^R_H(\omega)+\beta_{\omega_K\omega}^Rf^{R*}_H(\omega) \right] \;, \ee
with $\alpha$ and $\beta$ are given by\footnote{Their explicit values are given  in Eqs. (4.14b)-(4.14e) of Ref. \cite{paper2013}.}
\bea \alpha_{\omega_K,w}^L &=& \left(f_H^K(\omega_K, x), f_H^L(\omega)\right)  \\
\beta_{\omega_K,w}^L &=&\left(f_H^K(\omega_K, x), f_H^{L*}(\omega)\right) \\
\alpha_{\omega_K,w}^R &=&  \left(f_H^K(\omega_K, x), f_H^R(\omega)\right)  \\
\beta_{\omega_K,w}^R &=&  \left(f_H^K(\omega_K, x), f_H^{R*}(\omega)\right)\ . \eea
All scalar products are evaluated on the past horizon.

The exact expressions for the E.F. modes  $f_H^R$ can be obtained in the $R$ region by noting that the mode equation there
is exactly the same as the Schrodinger equation for scattering in one dimension off a step function potential.  To see this note that  we can write
\be f^R_H = \frac{e^{-i \omega t}}{\sqrt{4 \pi \w}}  \chi^R_H(x^*)  \;. \ee
Then the mode equation~\eqref{aet1} becomes
\be \frac{d^2 \chi^R_H}{d x^{* \; 2}} + [\omega^2 - m_R^2 \theta(x^* - x_0^*)] \chi^R_H = 0 \;. \label{chi-eq} \ee

Enforcing the continuity conditions at the barrier $x^*={}_Rx_0^*$ is very simple.  The $f_H^R$ modes are incident from the past horizon where they are given by~\eqref{rm}.
A mode propagates upstream in $R$ until the barrier at $x^*={}_Rx_0^*=0$ is reached.  If $\omega>m_R$ the mode is partially transmitted towards $x^*=+\infty$ in the form $\chi^R_H = A_\w e^{i\sqrt{\omega^2-m_R^2}x^*}$ as a massive $u$ mode\footnote{Although the modes are massive and so do not propagate along null geodesics, we use the ``u v'' notation to denote right moving and left moving waves respectively.} and partially reflected towards the horizon in the
massless form $D_\w e^{-i\omega x^*}$.  Continuity of the radial mode functions and their first derivatives at the barrier gives
\bes \bea A_\w &=& \frac{2 \omega}{\omega + k_R} \ , \\
D_\w &=& \frac{\omega - k_R}{\omega + k_R}  \ , \\
k_R &\equiv& \sqrt{\omega^2 - m_R^2}  \ .  \label{kR-def}
\eea \ees
Thus for $x > 0$
\bes \bea
f_H^R &=& \frac{1}{\sqrt{4\pi\omega}}e^{-i\omega t}\left[ e^{i\omega x^*}+ \frac{\omega - k_R}{\omega + k_R}e^{-i\omega x^*}\right] \,, \;\;\;  x^*< {}_Rx_0^* =0\;, \\
      &=&  \frac{1}{\sqrt{4\pi k_R}}  \left(  \frac{2\sqrt{k_R\omega}}{\omega + k_R} \right)  e^{-i\omega t}  e^{i k_Rx^*}  \,, \;\;\;  x^*> {}_Rx_0^*=0 \;.  \label{mnmn}
 \eea \ees
If $0\leq \omega <m_R$ there is total reflection (see also~\cite{Rousseaux2011}): the reflection coefficient for the mode is one and the mode is exponentially damped to the right of the barrier.

The reflected portions $\frac{D_\w}{\sqrt{4\pi\omega}} e^{-i\omega (t+x^*)}$ of the $f_H^R(\omega)$ modes enter the horizon and travel unaffected in the
interior region $L$ ($x<0$) until they encounter the step function at ${}_Lx_0^*$ where particle production occurs.\footnote{One could also think of this as scattering.  However in $L$, $x^*$ is the time coordinate so the scattering is anomalous because it involves negative norm modes (see for instance Sec. 9.6 of \cite{libro}).}
  By applying the same type of matching procedure as before and setting ${}_Lx_0^*=0$ we have in $L$
\bes \bea  f_H^R(\omega)&=& \frac{1}{\sqrt{4\pi\omega}} \left(\frac{\omega - k_R}{\omega + k_R}\right) e^{-i \omega t} e^{-i \omega x^*} , \; x^* < {}_Lx_0^*=0 , \\
                        &=& \frac{1}{\sqrt{4\pi k_L}}  \left(\sqrt{\frac{k_L}{\omega}} \frac{\omega-k_R}{\omega + k_R}\right) e^{-i\omega t} \left[ \frac{k_L+\omega}{2k_L} e^{-ik_Lx^*} + \frac{k_L-\omega}{2k_L}e^{ik_Lx^*}\right] , \; x^* > {}_Lx_0^*=0 ,\ \ \ \ \ \ \ \ \ \ \ \  \label{lulu} \\
                k_L   &\equiv& \sqrt{\omega^2+m_L^2} . \label{kL-def} \eea \ees
Note that $k_L$ is real for all values of $\omega$.

For the $f_H^L(\omega)$ modes one similarly obtains
\bes \bea f_H^L(\omega) &=& \frac{1}{\sqrt{4\pi\omega}} e^{i\omega t}e^{-i\omega x^*} \;, \qquad x^*<{}_Lx_0^*=0 \\
                  &=& \frac{1}{\sqrt{4\pi k_L}}e^{i\omega t}[ \sqrt{\frac{k_L}{\omega}} \frac{k_L+\omega}{2k_L} e^{-ik_Lx^*}+\sqrt{\frac{k_L}{\omega}} \frac{k_L-\omega}{2k_L}e^{ik_Lx^*}] \;,
     \qquad x^*>{}_Lx_0^*=0  \;.   \ \ \ \ \ \ \ \ \label{gigi} \eea \ees

The basis is completed by including the $f_I(\omega)$ modes coming from $x=+\infty$ and traveling downstream. These modes exist only for $\omega > m_R$
and originate as massive incoming $v$ modes.  Their normalization is discussed in Appendix~\ref{SCRIminusNomalization}.
In region $R$ ($x>0$) they are
\bes \bea  f_I(\omega)&=& \frac{1}{\sqrt{4\pi k_R}} e^{-i\omega t} [ \frac{k_R-\omega}{k_R+\omega}e^{ik_Rx^*} + e^{-ik_R x^*} ] \;, \qquad
   x^*>{}_Rx_0^*=0 \;, \label{zz} \\
   &=& \frac{1}{\sqrt{4\pi \omega}} \frac{2\sqrt{k_R\omega}}{k_R+\omega}e^{-i\omega t}e^{-i\omega x^*} \;, \qquad  x^*< {}_Rx_0^*=0  \;.   \label{uu} \eea \ees
The $f_I(\omega)$ modes maintain the form
of Eq. (\ref{uu}) in the $L$ ($x<0$)  region until they encounter the step discontinuity at $x^*={}_Lx_0^*=0$ where they undergo particle production.
Since the $L$ region is supersonic, both the $v$ mode and the $u$ mode travel downstream towards $x=-\infty$.
Noting that $x^{*}$ is a time coordinate in this region, the modes are given by the expressions
\bes \bea f_I(\omega)&=& \frac{1}{\sqrt{4\pi k_R}} \frac{2k_R}{k_R+\omega} e^{-i\omega t} e^{-i\omega x^*} , \; x^*<{}_Lx_0^* ,  \\
&=& \frac{1}{\sqrt{4\pi k_L}}  \frac{2\sqrt{k_Rk_L}}{k_R+\omega} e^{-i\omega t} \left[ \frac{k_L+\omega}{2k_L} e^{-ik_Lx^*} + \frac{k_L-\omega}{2k_L} e^{ik_Lx^*}\right], \;
  x^*>{}_Lx_0^* \; .  \label{mcmc}  \eea  \ees

Having found the complete set of modes that corresponds to the Unruh state, we turn next to the construction of the
symmetric two-point function for the field, also called the Hadamard function.  A general expression for it is
\be \label{tpt} G^{(1)}(x,x')= \langle 0| \{ \hat \phi (T,x), \hat \phi (T',x') \} | 0 \rangle \ . \ee
An explicit expression for the Unruh state in was derived in~\cite{paper2013}.  One first substitutes~\eqref{exp} into~\eqref{tpt} which gives
\be G^{(1)}(x,x')=I+J\ , \ee
with
\bes \bea  I &=&\int_0^\infty d\omega_K  \left[f_{K}^{H}(\omega_K, t, x)f_{K}^{H*}(\omega_K, t^\prime, x^{\prime})+ f_{K}^{H*}(\omega_K, t, x)f_{K}^{H}(\omega_K, t^\prime, x^{\prime}) \right]\ , \label{ii} \\
J &=& \int_0^\infty d\omega  \left[f_I(\omega, t, x)f_I^{*}(\omega, t^\prime, x^{\prime})+ f_I^{*}(\omega, t, x)f_I(\omega, t^\prime, x^{\prime}) \right] \ . \label{jj} \eea \ees
Expressing the $f_H^K$ modes in terms of the $f_H^{L,R}(\omega)$ ones by means of the Bogoliubov transformation (\ref{expu}) one finds that $I$ can be rewritten as
\bea I= \int_0^\infty d\omega \frac{1}{sinh \left(\frac{\pi \omega}{\kappa}\right)}  \left[f_{L}^{H}(\omega, t, x)f_{R}^{H}(\omega, t^\prime, x^{\prime})+\right. f_{L}^{H*}(\omega, t, x)f_{R}^{H*}(\omega, t^\prime, x^{\prime}) \nonumber \\   +f_{R}^{H}(\omega, t, x)f_{L}^{H}(\omega, t^\prime, x^{\prime})+f_{R}^{H*}(\omega, t, x)f_{L}^{H*}(\omega, t^\prime, x^{\prime}) \nonumber \\
 +cosh\left(\frac{\pi \omega}{\kappa}\right)\left[f_{L}^{H}(\omega, t, x)f_{L}^{H*}(\omega, t^\prime, x^{\prime})+f_{L}^{H*}(\omega, t, x)f_{L}^{H}(\omega, t^\prime, x^{\prime})\right. \nonumber \\ \quad \left. \left.+f_{R}^{H}(\omega, t, x)f_{R}^{H*}(\omega, t^\prime, x^{\prime})+f_{R}^{H*}(\omega, t, x)f_{R}^{H}(\omega, t^\prime, x^{\prime})\right]\right] \ . \label{zuz} \eea

It is important to note that while our toy model is simple enough that analytic expressions have been obtained for all of the mode functions, the resulting integrands in the
above expressions are complicated enough that we have computed them numerically rather than analytically.  Since the points are split no renormalization counterterms are necessary and because the field has a mass, there are no infrared divergences as there are in the massless case.

\section{Results\label{sec:results}}

Our main result is that the existence of a mass term in the mode equation, however small it may be, fundamentally changes the nature of the solutions, leading to
a much more complex structure than occurs in the massless case. To provide some context for this structure we first investigate the behavior of the two point function in the Minkowski vacuum state in two dimensions for both massless and massive scalar fields.  Then we review the massless case for both our two dimensional analogue BH model and the two dimensional model in~\cite{paper2013} which is obtained via dimensional reduction.  The primary difference between the two is the existence of an effective
potential in the latter case resulting from the dimensional reduction.  This effective potential causes scattering of the mode functions and particle production to occur.
After that we discuss the results for our model and finish with a discussion of what happens as the mass becomes very small.

\subsection{Correlation function in flat space}
\label{sec:flat}

To gain insight into the features seen in the black hole analogue it is useful to investigate the behavior of the two-point correlation function in flat 2D spacetime in both the massless and massive cases.  The solutions to the mode equations which correspond to the Minkowski vacuum are
\bea  f_k &=& \frac{1}{\sqrt{4 \pi \omega}} e^{-i \omega t + i k x} \;, \label{mode-flat}  \\
      \omega &=& \sqrt{k^2 + m^2}  \;.  \nonumber \eea
The two point function is
\bes \bea  \la \{ \hat{\phi}(t,x),\hat{\phi}(t',x') \} \ra &=& \frac{1}{4 \pi} \int_{-\infty}^\infty \frac{dk}{\omega} \left ( e^{-i \w (t - t')+ i k (x-x')} +  e^{i \w (t - t')- i k (x-x')} \right)  \label{2pt-flat-1a} \\
    &=& \frac{1}{\pi} \int_0^\infty \frac{dk}{\w} \cos[\w (t-t')] \cos[k (x-x')]  \;, \label{2pt-flat-1b} \eea \ees

In the massless case $\w = k$ and there is an infrared divergence which can be removed by imposing a lower limit cutoff $\lambda$.  Using
the identity
\be \cos[k (t-t')] \cos[k (x-x')] = \frac{1}{2} \left( \cos[k(|t-t'|+|x-x'|)] + \cos[k(|t-t'|-|x-x'|)]  \right)   \;, \ee
it is easy to show that
\be  \la \{ \hat{\phi}(t,x),\hat{\phi}(t',x') \} \ra = -\frac{1}{2 \pi} \left( {\rm ci}[\lambda (|t-t'|+|x-x'|)] + {\rm ci}[\lambda(|t-t'|-|x-x'|)] \right)\;, \label{2pt-flat-m-0} \ee
where ${\rm ci}(x)$ is a cosine integral.  For large enough separations of the points there will be structure in the two-point correlation function but it depends on the cutoff $\lambda$.  Further if one or more derivatives with respect to the spacetime coordinates are taken, as they would be for the density density correlation function, then the
infrared divergence disappears so one can take the limit $\lambda \to 0$.  It is easy to see that in this case all of the structure disappears except for the peak that occurs when the points come together and the peaks that occur when $|x-x'| = |t-t'|$.

The massive case is completely different.  Computing the integral in~\eqref{2pt-flat-1b} gives
\bes \bea   \la \{ \hat{\phi}(t,x),\hat{\phi}(t',x') \} \ra &=& -\frac{1}{2} N_0(m \sqrt{(t-t')^2-(x-x')^2})\;, \qquad |t-t'| > |x - x'| \;, \label{2pt-flat-N0}\\
         &=& \frac{1}{\pi} K_0(m \sqrt{(x-x')^2-(t-t')^2}) \;, \qquad |x-x'| > |t - t'|  \;, \label{2pt-flat-K0} \eea \label{2pt-flat-2} \ees
where $N_0$ and $K_0$ are Bessel functions.  If $|t-t'| > |x - x'|$ then there is a series of parallel correlation peaks which occur because $N_0$ is an oscillating function
with the amplitude of the oscillations damped as its argument gets larger in magnitude.  But if $|x-x'| > |t - t'|$ then the only correlation peak is the one where the
points come together.  The boundary between the two regions, $|t-t'| = |x - x'|$, is also a correlation peak in both the massless and massive cases.  For the analogue
black hole, evidence that the correlation peaks in the massless case correspond to boundaries in the massive case when both points are inside the horizon is given below.

One can find the locations of the peaks along with the boundary between where multiple peaks do and do not occur using the method of stationary phase on the integral in~\eqref{2pt-flat-1a}.  It is easily seen that there are stationary phase points for real values of $k$ only if $|t-t'| > |x - x'|$.  Further, if stationary phase points
do exist, one can use the method to reproduce to leading order in $(t-t')^2 - (x-x')^2$ the result in~\eqref{2pt-flat-N0}.

\begin{figure}[h]
\includegraphics[width=3in]{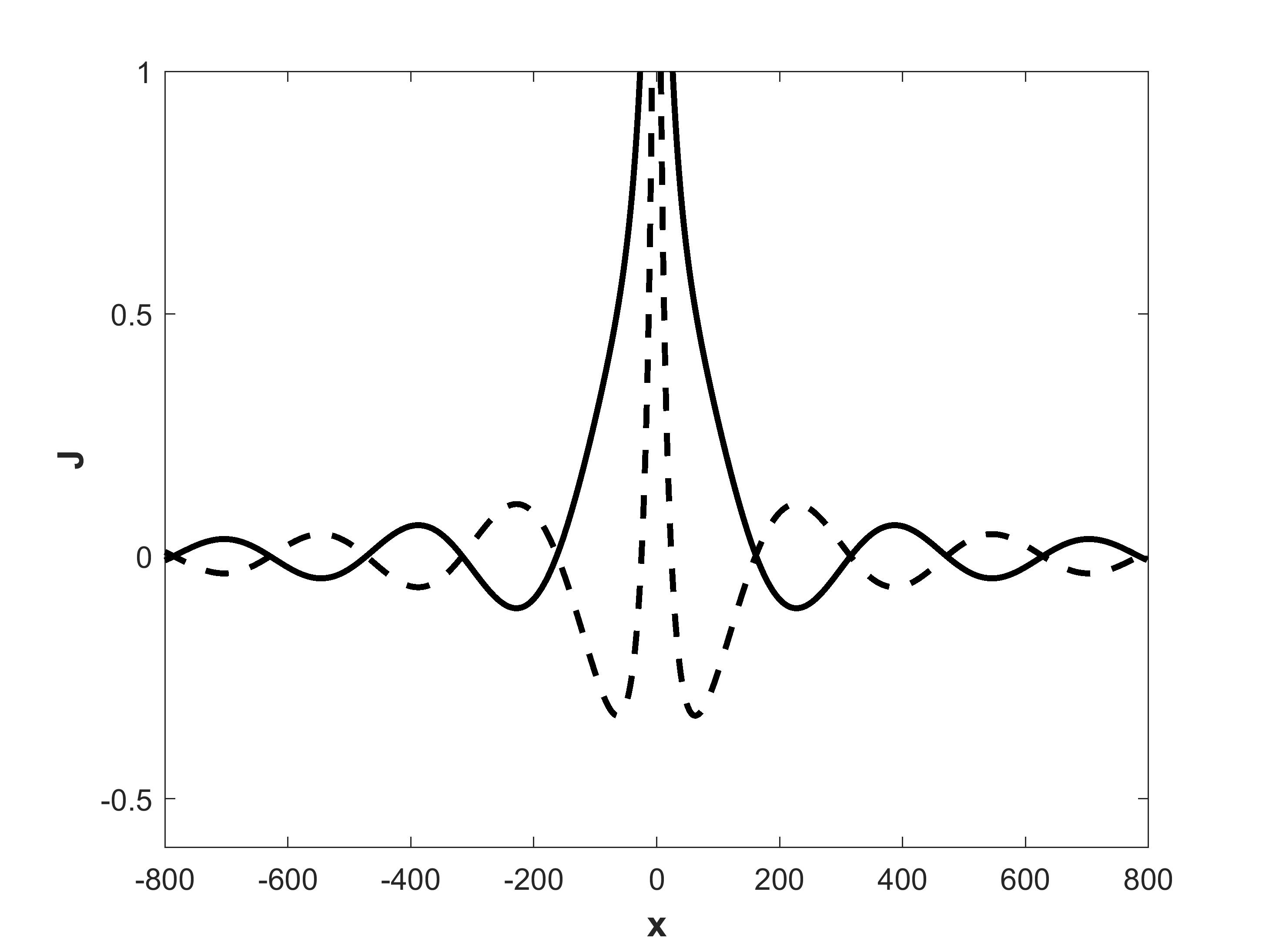}
\caption{\label{Fig:LeftandRighMoving}  The real part of solutions for $J_{(left-moving)}$(dashed) and $J_{(right-moving)}$(solid).
Note that when added together there is clear cancellation of the smaller peaks and only the single large peak that occurs when the points
come together survives.}
\end{figure}

When $|t-t'| < |x - x'|$ there is no series of parallel peaks as can be seen from Eq.~\eqref{2pt-flat-K0}.  However, if one breaks up the contribution to the two point function
in this case into right and left moving modes and considers them separately then one finds a series of peaks that are parallel to the main peak when the points come
together.  The integrals in question are
\bea J_{(left-moving)} &=& \frac{1}{4 \pi} \int_{0}^\infty \frac{dk}{\omega}  e^{-i \w (t - t')- i k (x-x')} \;, \nonumber \\
  J_{(right-moving)} &=&  \frac{1}{4 \pi} \int_{-\infty}^0 \frac{dk}{\omega}  e^{-i \w (t - t')- i k (x-x')} \;, \label{J1LR} \eea
In Fig.~\ref{Fig:LeftandRighMoving} the real parts of the solutions are plotted for the case that $t-t' = \frac{1}{2} (x-x')$.
It is clear that what happens is that cancellations cause the structure to disappear when these integrals are added together.
Similar effects occur in the BH analogue model.  In this case the integrands are slightly different so the contributions from the initially left moving
$f^{R,L}_H$ modes to these peaks is not completely cancelled by the contribution from the initially right moving $f_I$ modes. However, the cancellation is good enough that the remaining structure is smaller by at least three orders of magnitude than the other
more prominent structures discussed below.

Since multiple correlation peaks occur for any $m > 0$, in some sense there is an abrupt transition between the two point function for a massless and a massive field.
However, it is instructive to see what happens as the mass decreases in the region where there are multiple peaks.
Because of the factor of $m^2$ in the Hankel function in~\eqref{2pt-flat-N0} the separations between the peaks gets larger as $m$ gets smaller.  Thus if one restricts attention to a finite region near the origin, then one finds that as the mass decreases the structure effectively moves out of this region.  As discussed below, we find something similar in the BH analogue case.

\subsection{Review of the massless BH analogue case}

Before discussing the structure of the two point function in the massive case it is useful to review what happens in the massless case for a BH analogue, i.e. when one sets $m=0$ in the mode equation
~\eqref{kgt}. This case was investigated in the pioneering work~\cite{paper1}.  There it was found that, along with the usual peak which occurs when the points come together, the density density correlation function in a BEC (which is simply related to $G(x,x')$, see ~\cite{paper1}) has a peak when one point is inside and one point is outside the horizon, revealing
the correlations between the Hawking particles and the negative energy partners inside. This is the smoking gun of the Hawking effect which has been found in the experiments of Steinhauer \cite{jeff,jeff2}.

\begin{figure}[h]
\includegraphics[width=3in]{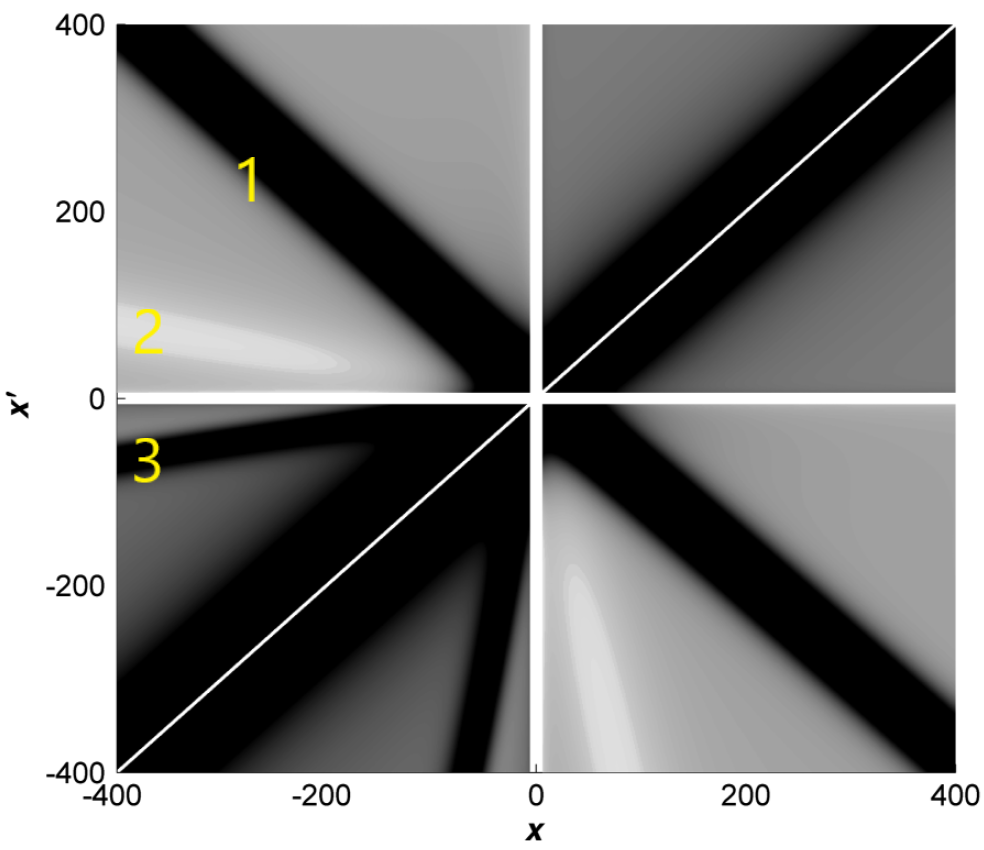}
\caption{\label{Fig:2013PRD BEC DD Plot} Density-density correlation function for 1+1D BEC BH analog with no masslike term 
which shows 3 peaks.  The figure is reproduced from~\cite{paper2013} with the numbers superposed on that figure.   }
\end{figure}

\begin{figure}[h]
\includegraphics[width=4in]{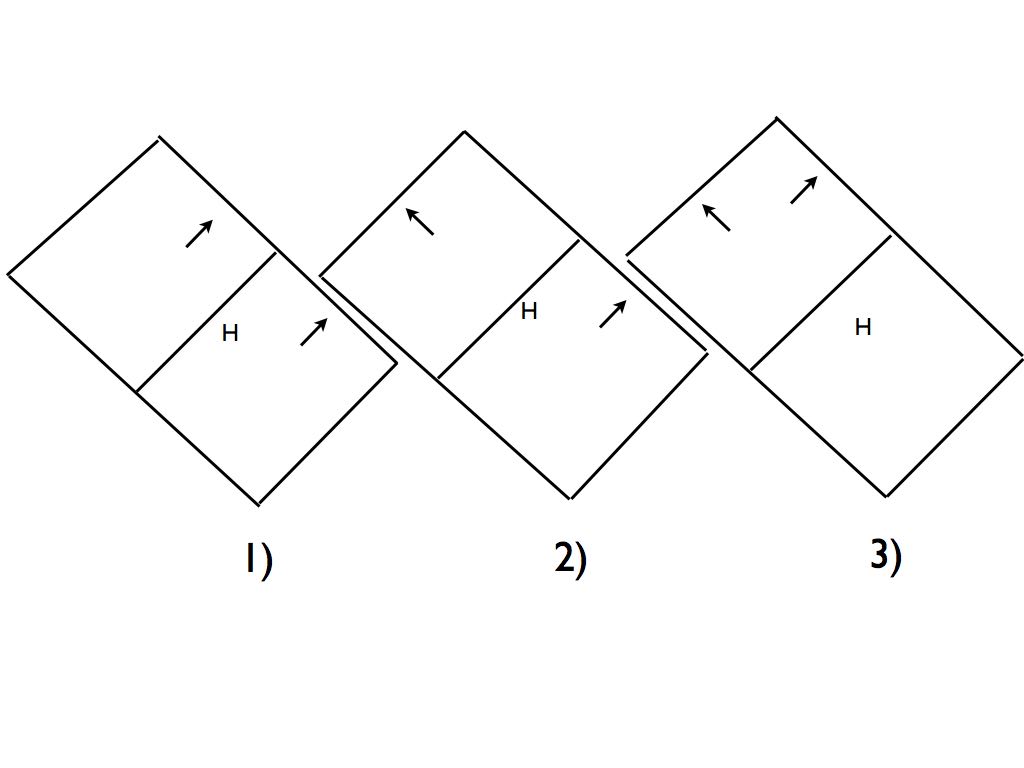}
\caption{\label{three-diagrams} Correlations between the modes found in the massless case.}
\end{figure}

Numerical studies using a full quantum treatment of the BEC~\cite{paper2} and studies using QFT in curved space techniques~\cite{paper2013}
confirmed the existence of the in-out peak and showed the existence of two others.\footnote{The reason the extra peaks show up in the treatment of~\cite{paper2013} is that
the effective potential which results from a dimensional reduction from four to two dimensions is included and this results in scattering and particle production effects which do not occur when the effective potential is not present. }
  One was a weaker,  in-out peak and the other a correlation peak that occurs when both points are inside the horizon.  Figure \ref{Fig:2013PRD BEC DD Plot} contains a plot from Ref.~\cite{paper2013} which shows these peaks.   We have superposed the numbers 1, 2, and 3 on this plot to label the main in-out peak, the secondary in-out peak, and the in-in peak respectively.
Penrose diagrams which sketch the corresponding correlations between the modes for the three peaks are given in Fig \ref{three-diagrams}.
One sees that peak (1) is of the $u-u$ type whereas peaks (2) and (3) are of the type $u-v$ and require for their existence a scattering of the massless modes.

\subsection{Primary results for the massive case}

\begin{figure}
\includegraphics[width=5.in]{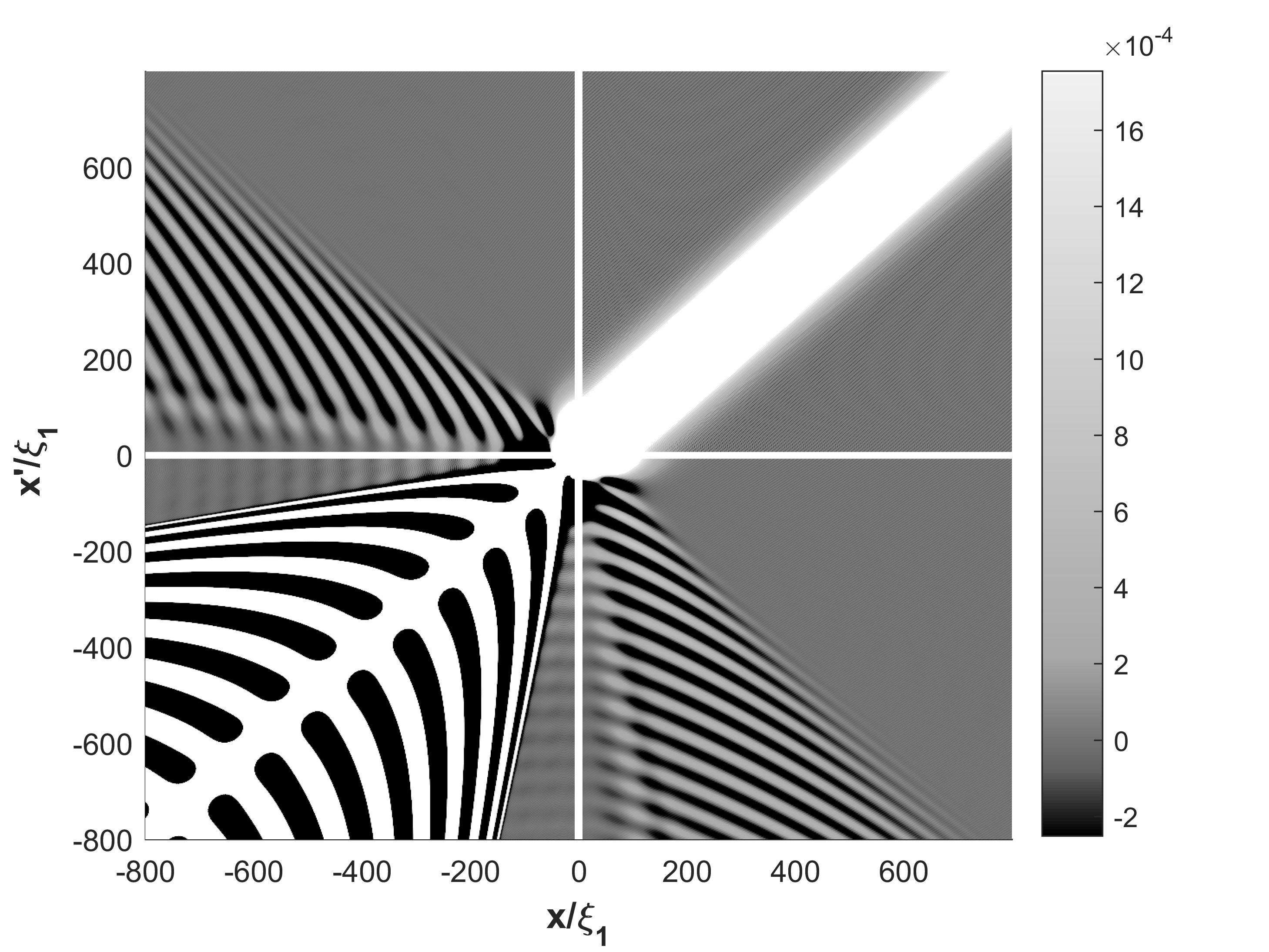}
\caption{\label{Fig:ExtremeRange} Plot for
the two point function for $m = 4 \times 10^{-2}m_a$.  The horizontal and vertical white stripes correspond to regions near the
horizon which are excluded from the plot.  The contrast has been arranged to show the primary structures that have been observed in this case.  As a result regions which appear
to be pure white aside from the exclusions are regions where the two point function is larger than $1.7 \times 10^{-3}$.  For the pure black regions the two point function is smaller than $-2\times 10^{-4}$.
   }
\end{figure}
The most prominent structures we have found for the two-point correlation function are shown in  Fig.~\ref{Fig:ExtremeRange}.  In each quadrant there is also a group
of correlation peaks which are so small that they do not show up on the scale of the plot.  Because they are insignificant compared to the more prominent peaks we do not discuss them further in this subsection but they are briefly discussed in Sec.~\ref{sec:stationary} below.

\begin{figure}
\includegraphics[width=5.in]{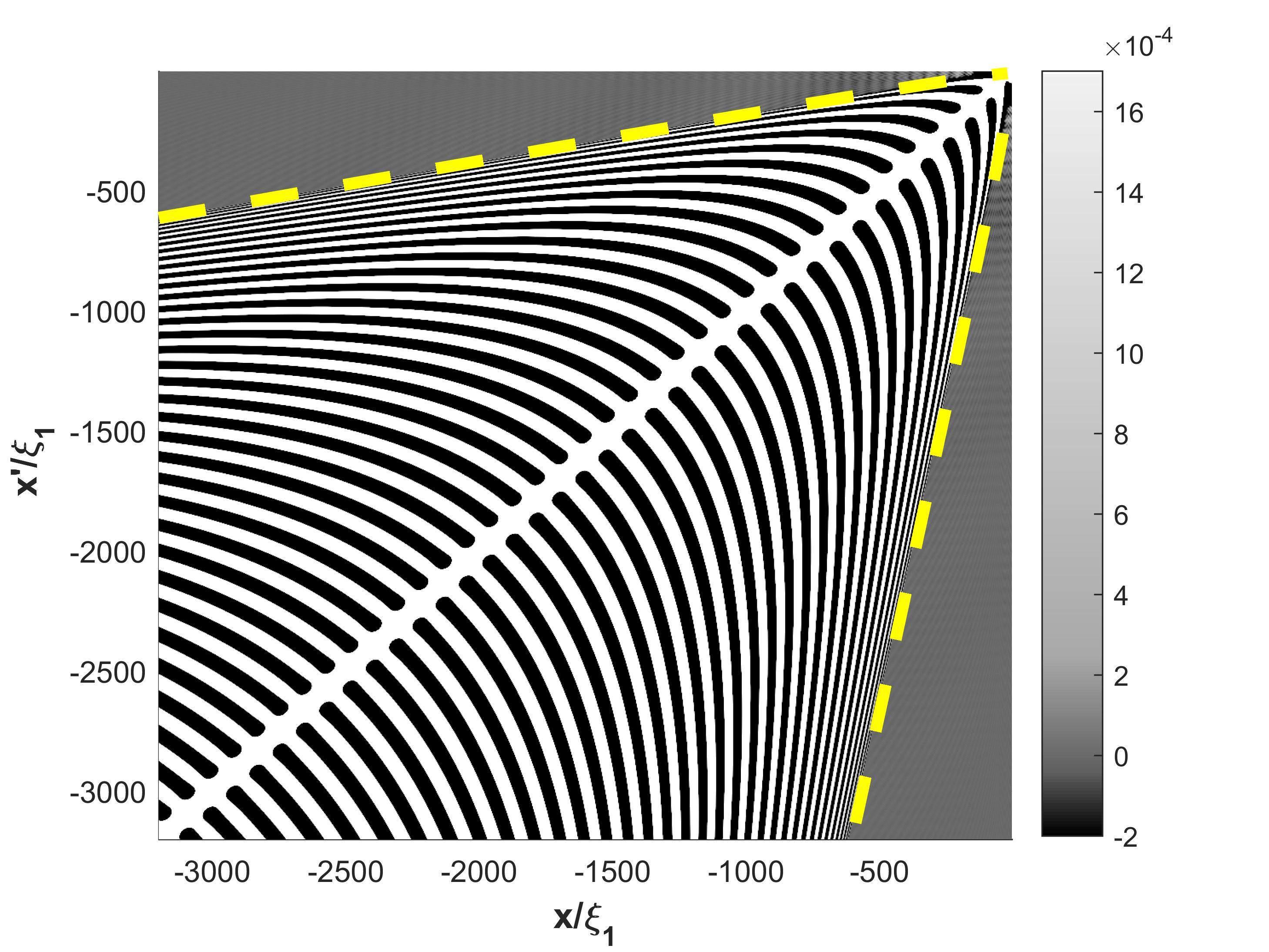}
\caption{\label{Fig:in-in-detail}Plot of the two point function for $m=4\times 10^{-4}m_a$. The region shown is the lower left quadrant of Fig.~\ref{Fig:ExtremeRange}, where both points are in the interior of the analogue BH. Here the range of the plot has been extended to better illustrate the features in this region.  The dashed lines indicate the correlation peaks in the massless case with locations specified by~\eqref{m0-3}.  Note that the less pronounced set of peaks that is clearly visible in Fig.~\ref{Fig:ExtremeRange} is only visible here near the upper right corner of the plot.  The reason is that the amplitudes of these peaks decrease as the
separation between the points $x$ and $x'$ increases.  }
\end{figure}

\begin{figure}
\includegraphics[width=5.in]{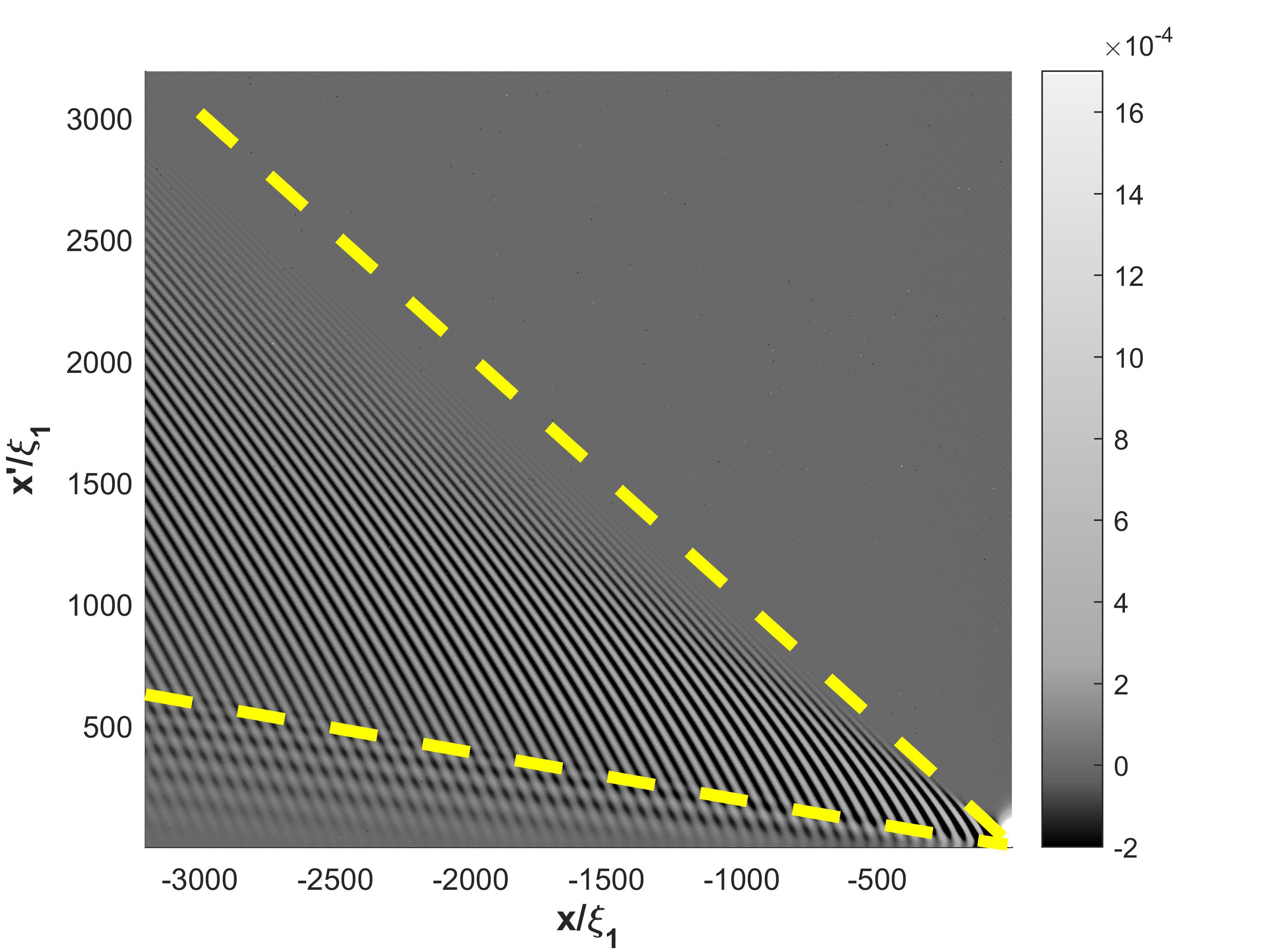}
\caption{\label{Fig:in-out-detail}Plot of the two point function for $m=4\times 10^{-4}m_a$. The region shown is the upper left quadrant of Fig.~\ref{Fig:ExtremeRange}, where one point is in the interior and the other point is in the exterior of the analogue BH. Here the range of the plot has been extended to better illustrate the features in this region.  The dashed lines indicate the locations of the correlation peaks in the massless case with locations specified by~\eqref{m0-1} and~\eqref{m0-2}.  }
\end{figure}

In contrast with the massless case, the structures in the massive case generally consist of groups of roughly parallel peaks which are sometimes separated by boundaries.  The locations of the boundaries when both points are inside the horizon correspond to the locations of the peaks that would occur in our simple model in the massless case\footnote{We show the peaks here in the case that there is no potential so that solutions to the radial mode equation in the massless case are linear combinations of  $e^{i\w x^*}$ and $e^{-i \w x^*}$.  The solutions are more complicated if there is a potential and the locations of the peaks may be altered somewhat.  Of course strictly speaking in the massless case there is no scattering or particle production if there is no potential so peaks (2) and (3) do not occur.  However, they would occur if there was scattering, although their locations might be
somewhat different.}.  When one point is inside and one point is outside the horizon the boundaries appear to be close to the location of the peaks in the massless case.
By examining~\eqref{jj} and~\eqref{zuz} it is not hard to show that for our simple model  peak 1) occurs when the contributions to the phase of a right moving mode in $R$
and a right moving mode in $L$  cancel.  For point $x$ in the $R$ region and $x'$ in the $L$ region this results in the condition
\be  (t-t') - (x^* - x^{*\,'}) = 0 \;. \label{m0-1} \ee
Similarly peak 2) occurs when the contributions to the phase of a right moving mode in $R$ and a left moving mode in $L$ cancels.  For point $x$ in the $R$ region and $x'$ in the $L$ region this corresponds to
\be (t-t') - (x^* + x^{*\,'}) = 0 \;.  \label{m0-2}  \ee
Peak 3) occurs when the contributions to the phase of a left moving mode in $L$ and a right moving mode in $L$ cancel.
 In some cases this occurs when
\bes \be (t-t') + (x^*+x^{*\;'})  = 0 \;, \label{m0-3}  \ee
while in others it occurs when
  \be (t-t') - (x^*+x^{*\;'})  = 0 \;. \label{m0-4}  \ee  \ees
Using~\eqref{t-T} with $T'=T$ and~\eqref{xstar} one can find the relationships between $x$ and $x'$ for these peaks.  These are plotted as the
dashed lines in Figs.~\ref{Fig:ExtremeRange},~\ref{Fig:in-in-detail}, and~\ref{Fig:in-out-detail}.

Arguably the most interesting part of Fig.~\ref{Fig:ExtremeRange} is the lower left quadrant where both points are inside the horizon.  This is where the most prominent peaks occur and it is also the location of the undulations which are discussed in the next subsection.
When both points are inside the horizon, the most prominent peaks have a well defined boundary. This boundary is at the location of the peaks that occur in the massless case when both points are inside the horizon.  The details of this region along with the locations of the two peaks in the massless case are shown in Fig.~\ref{Fig:in-in-detail}.

The two regions in Fig.~\ref{Fig:ExtremeRange} for which one point is inside and one point is outside the horizon are the top left and bottom right quadrants.  They
 also have a rich structure.  There are two groups of peaks, partially superposed, which are both less prominent than the group of large peaks which occur when both points are inside
 the horizon.   One group consists of a series of parallel peaks which are roughly diagonal.  In both quadrants the upper boundary of these peaks is approximately, but not exactly, where peak (1) occurs in the massless case, see Eq.~\eqref{m0-1} and Fig.~\ref{Fig:2013PRD BEC DD Plot}.  The other group consists
 of a series of peaks that are roughly parallel to the horizon.  The upper boundary of these peaks is in the approximate location of peak (2) in the massless case, see
 Eq.~\eqref{m0-2} and Fig.~\ref{Fig:2013PRD BEC DD Plot}.
  There is an overlapping region where both groups of peaks are superposed.
  The details of this region along with the locations of the two peaks in the massless case are shown in Fig.~\ref{Fig:in-out-detail}.

 The top right quadrant in Fig.~\ref{Fig:ExtremeRange} is the region where both points are outside of the acoustic horizon.  The only peak visible in the plots is the correlation peak that occurs when the points come together.  As mentioned above, in each quadrant including this one, there is at least one group of very small peaks that do not show up
 on the scale of the plot.

\subsection{Undulations}
\label{sec:undulations}

As discussed in the Introduction, undulations are expected to occur in the region where both points are far enough inside the horizon that the particle creation event has occurred.  The undulations were predicted in~\cite{mbh} for 2D spacetimes with an event horizon when a mass term is present in the mode equation.  It was shown there that if one uses only the low frequency modes to compute the two-point function then this function will be proportianal to ${cos(p_u x)cos(p_u x^\prime)}$, with ${p_u\propto m}$.  This leads to a cross hatched pattern with a periodicity that is proportional to the mass.

\begin{figure}
\includegraphics[width=3.5in]{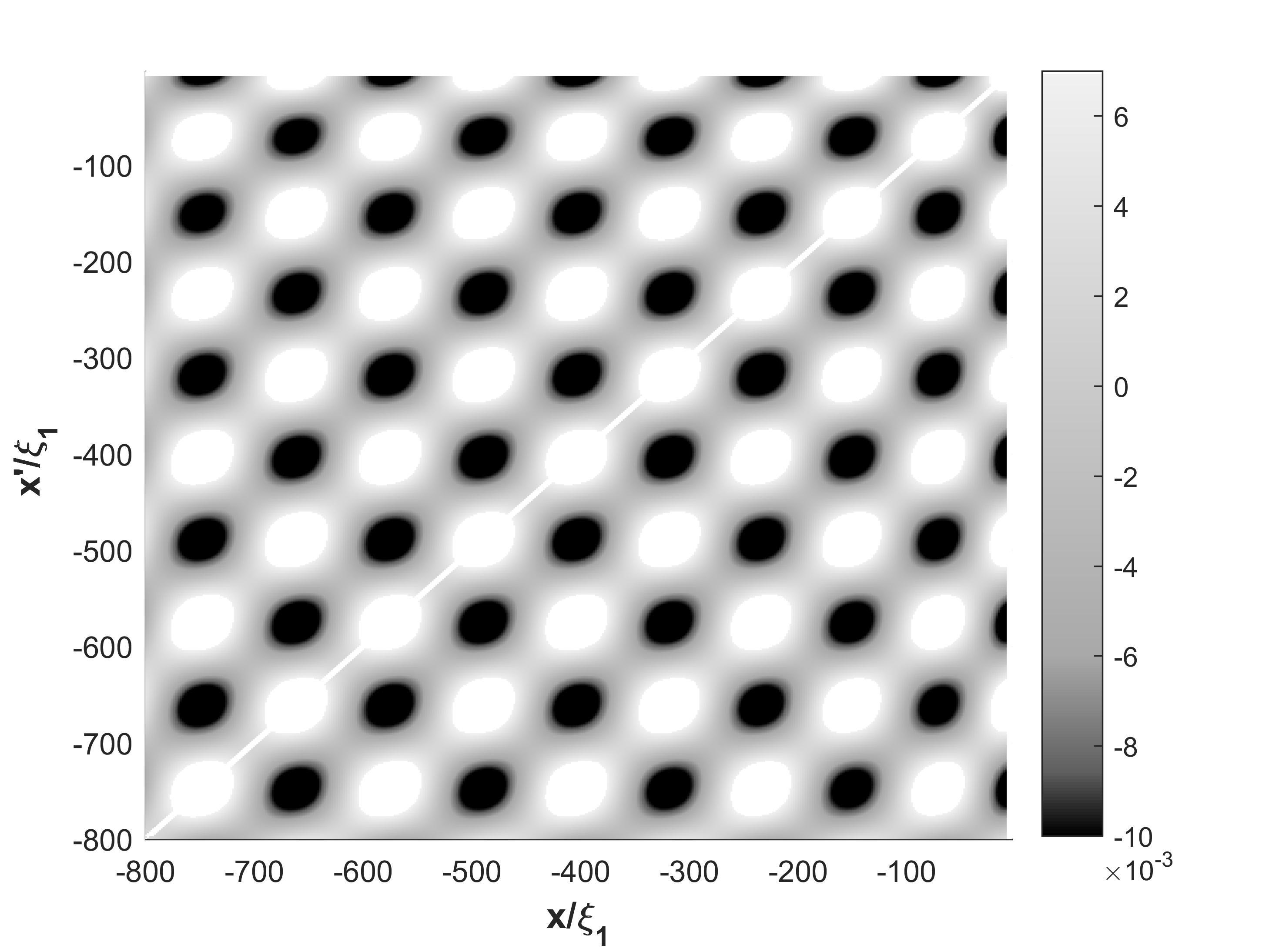}
\includegraphics[width=3.5in]{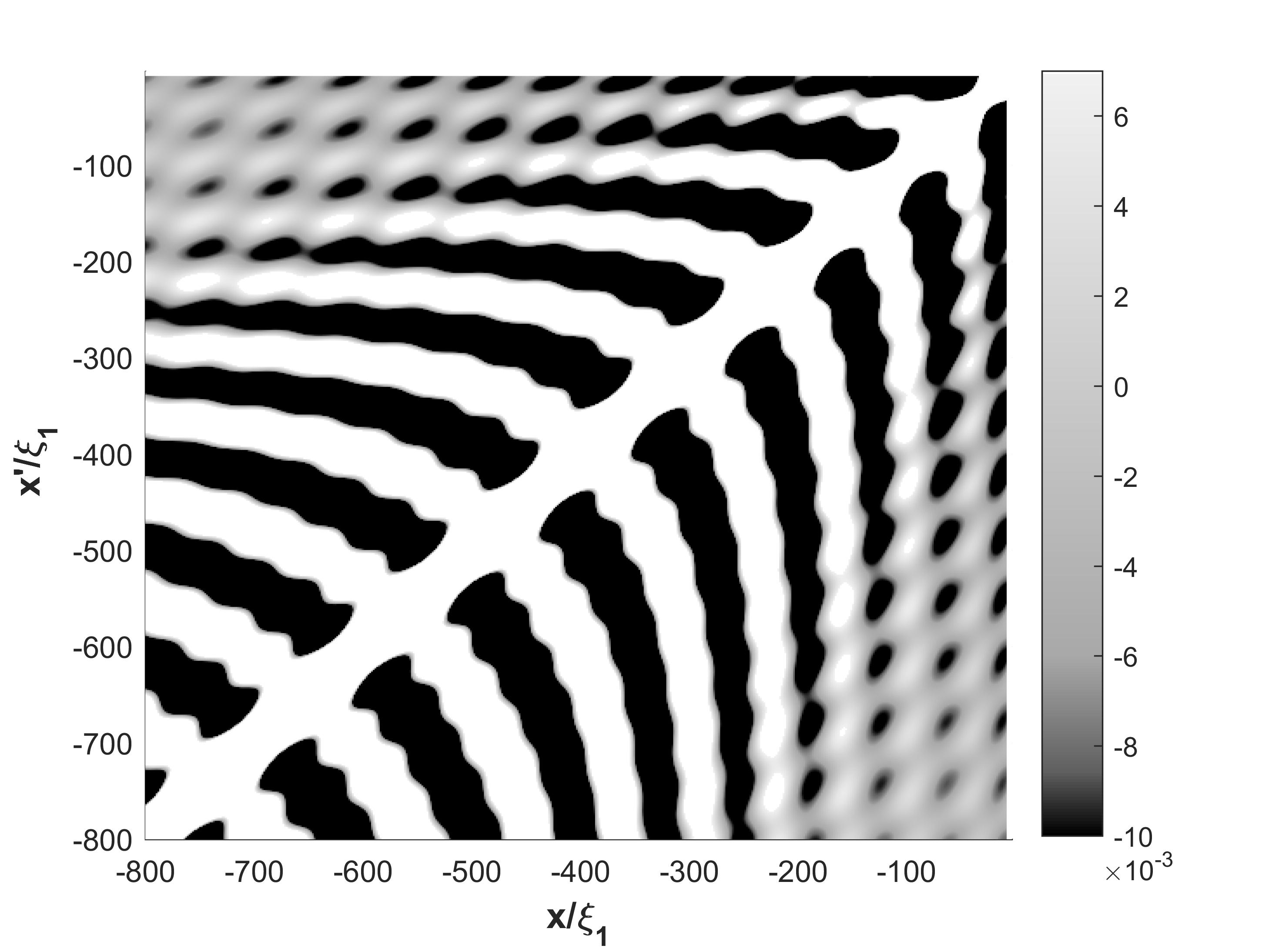}
\includegraphics[width=3.5in]{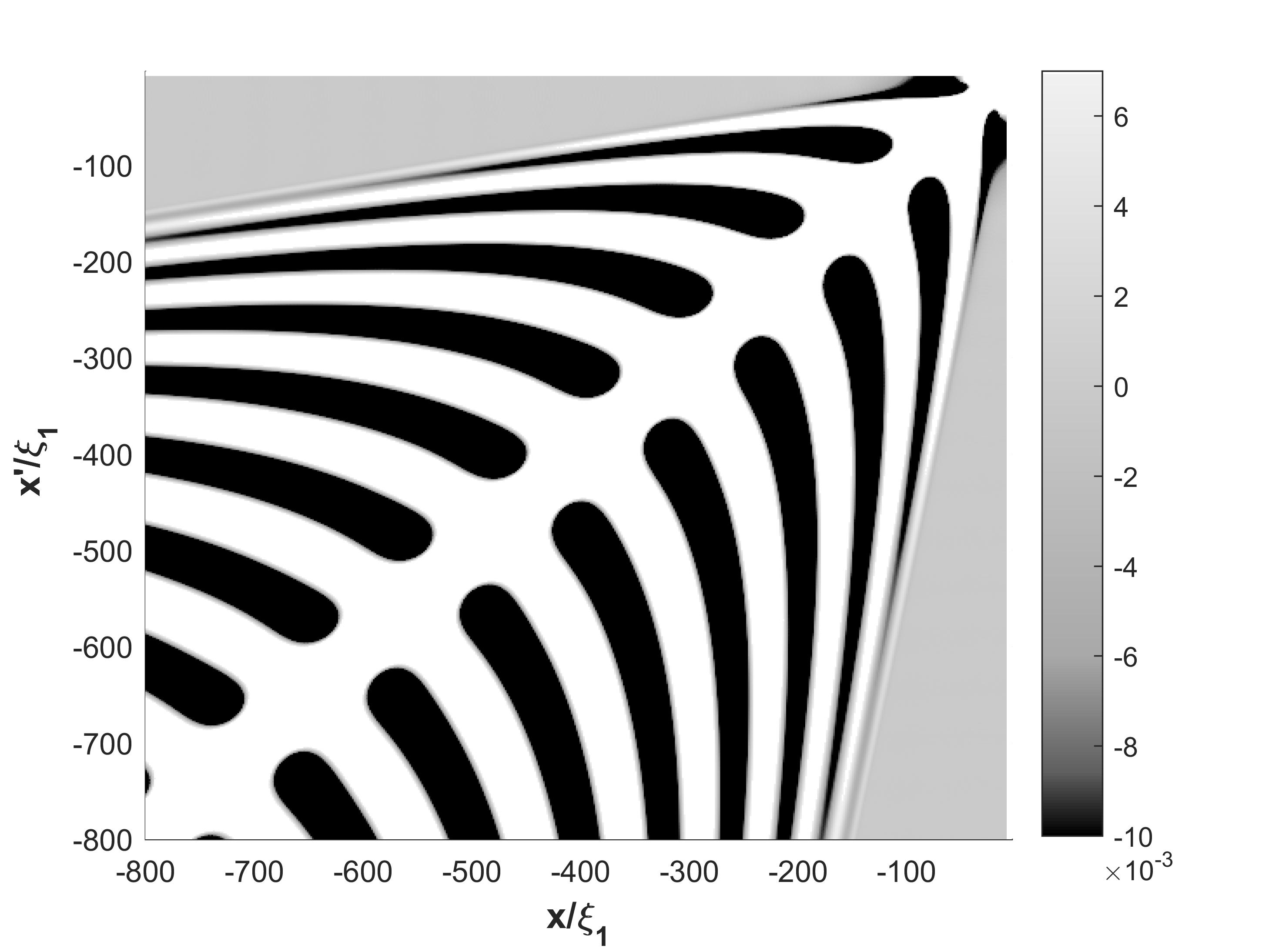}
\caption{\label{Fig:FrequencyContributions} Plots of the two point function for the case ${m=4 \times 10^{-2}m_a}$ for various frequency bands.   Top: Low frequency contribution to the two-point function for ${0\le\omega \le \frac{m_R}{100}}$ . Middle: Contributions from frequencies in the range ${0\le\omega \le \frac{m_R}{10}}$.  Bottom: Contributions from all frequencies.  Note that the scale of the color bar is larger than that in Fig. 6 which causes subtle differences in the third plot with the lower left quadrant of Fig. 6.  }
\end{figure}

It is clear by inspection of the lower left quadrant of Fig.~\ref{Fig:ExtremeRange} that no such pattern occurs.  To investigate the question of whether undulations are present
we show in Fig.~\ref{Fig:FrequencyContributions} the patterns obtained when certai integrals in~\eqref{jj} and~\eqref{zuz} are computed for certain
frequency ranges.  The top plots are only for the contributions from the low frequency modes, $0 \le \omega \le \frac{m_R}{100}$, and they reproduce the expected cross hatched pattern.  Thus the undulations are there.  The middle plots include
all modes with $0 \le \omega \le \frac{ m_R}{10}$ and their structure is essentially the same as that for the bottom plots which include all values of $\omega$.
Thus, when compared to the prediction in~\cite{mbh}, we find that the low frequency contribution does not dominate the total two point function and instead a much more complicated relationship is found for the total two point function.\footnote{We must note, however, that in our idealized model scattering coefficients depend only on $\omega,\ m_R,\ m_L$ and not on the surface gravity $\kappa$. In more realistic models those scattering coefficients that do not go to $1$ as $\omega\to +\infty$ will typically go to $0$ more rapidly than those in this model.  This could affect the results for the total two-point function presented here.}

\subsection{Stationary phase approximation}
\label{sec:stationary}

The stationary phase approximation can be used to gain some insight into the primary group of peaks which occur in the two point function when both points are
inside the horizon.  Not all terms in the integrand in this case have stationary phase points for positive real values of $\w$, but those with exponentials
whose arguments are of the forms
\bes \bea   && \mp i \w (t-t')  \pm i k_L (x^* + x^{*'})  \\
            &&  \mp i \w (t-t')  \mp i k_L (x^* + x^{*'})    \;, \eea \ees
have stationary phase points at
\be \w = \frac{m_L (t-t')}{\sqrt{(x^*+x^{*\,'})^2 - (t-t')^2}} \;, \label{stat}
 \ee
if $|x^* + x^{*\,'}| > |t-t'|$ and $t-t'>0$.  Thus the boundary between where a stationary phase point exists for positive values of $\w$ and where
one does not is given by the condition $|x^* +x^{*\,'}| = |t-t'|$.
This is the condition~\eqref{m0-4}  which corresponds to the location of correlation peak (3) in Fig.~\ref{Fig:2013PRD BEC DD Plot}.

\begin{figure}[h]
\includegraphics[width=3in]{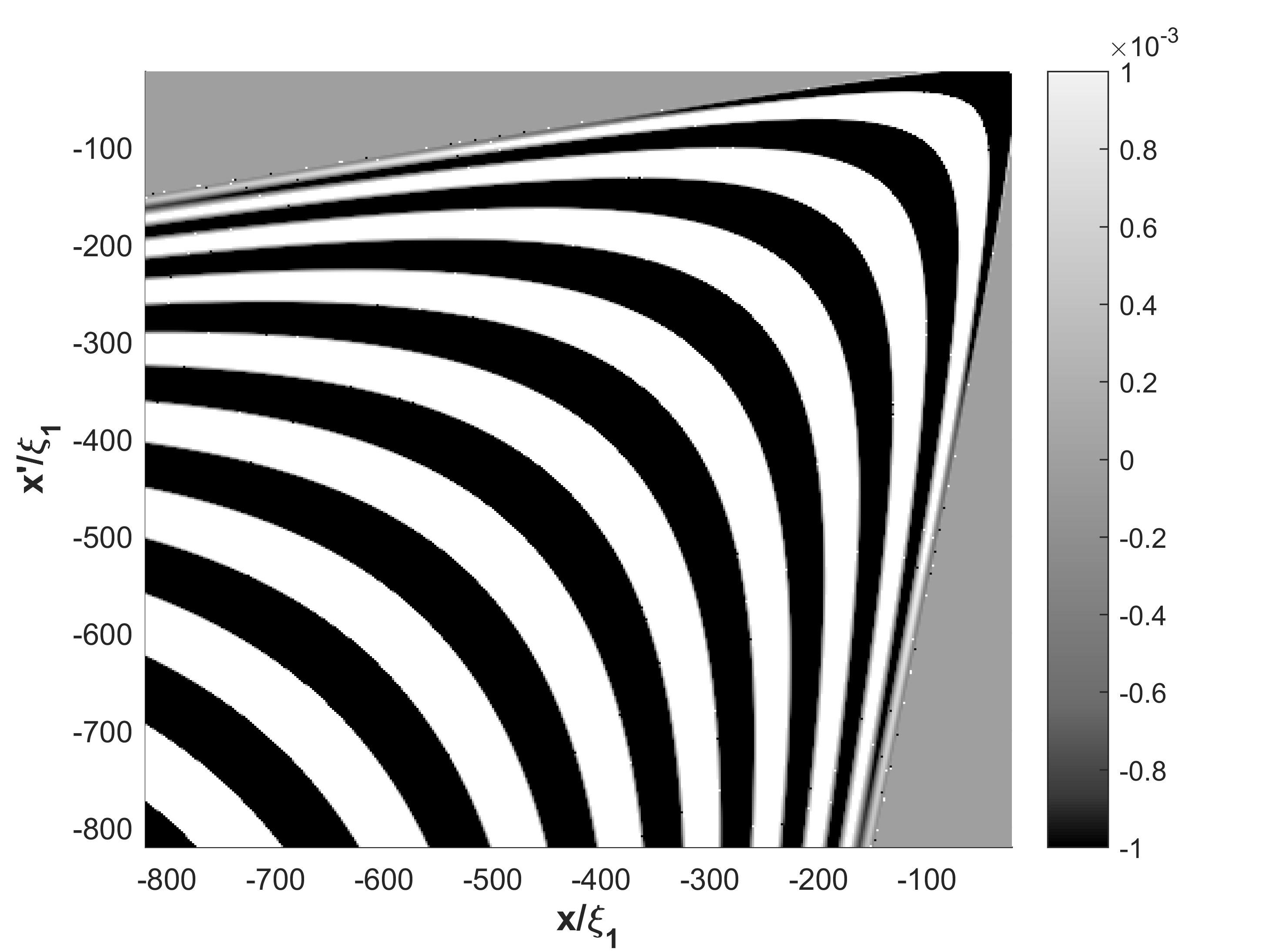}
\caption{\label{fig:stationary} Stationary phase result for part of the contribution to the two-point correlation function when both points are inside the horizon from products of the $f_h^L$ modes. One contribution consist of an upstream mode at point $x$ and a downstream mode at point $x^\prime$. The other has the point $x$ and $x^\prime$ for these modes reversed. }
\end{figure}

The unusual structure of the most prominent group of peaks when both points are inside the horizon, along with their boundaries, can be reproduced
by the stationary phase approximation for some of the integrands which contribute to the two point function.\footnote{The range of values of $x$ and $x'$ for which
there is a stationary phase approximation is more limited for the $f_I$ modes than is given by the conditions~\eqref{stat} because the minimum value of $\w$ for those
modes is $m_R$.  When the integrals are computed exactly
there are partial cancellations which occur between the contributions from the $f_H$ and $f_I$ modes.  Because of the extra limitation in the range of the stationary phase approximation for the $f_I$ modes, there are ranges of values of $x$ and $x'$ for which such cancellations cannot occur for the stationary phase approximation.  As a result the structure that appears due to the full stationary phase approximation looks significantly different from that which is due to the full numerical calculations in these regions.}
The stationary phase approximation for this case is plotted in Fig.~\ref{fig:stationary}.
 Comparison with
Fig.~\ref{Fig:in-in-detail} shows that the large peaks are reproduced by the approximation along with their boundaries.
The other group of peaks in Fig.~\ref{Fig:in-in-detail}  are not reproduced by the stationary phase approximation.

There is no stationary phase approximation for the integrals when both points are outside the horizon and far from it for the values of $t-t'$ that we are using.
Recall these are obtained from~\eqref{t-T} with the condition $T = T'$.  This is consistent with the lack of prominent correlation peaks in this case as shown in Fig.~\ref{Fig:ExtremeRange} where the only peak visible is that which occurs when both points come together.

When one point is inside and one point is outside the horizon the stationary phase approximation reproduces the group of peaks which are approximately parallel to the
horizon.  The other group of peaks which are approximately diagonal are not reproduced by this approximation.

As mentioned previously there is some structure in each case, in-in, in-out, and out-out that is too small to show up on the plots.  The contributions to this structure
come from terms in the mode integrals for which there is no stationary phase approximation in the integration range.  In most cases we find that there is a partial
cancellation between the contributions from the $f^{R,L}_H$ and the $f_I$ modes.

\subsection{Behavior as $m \rightarrow 0$ \label{sec:mgoestozero}}

\begin{figure}[h]
\includegraphics[width=2.7in]{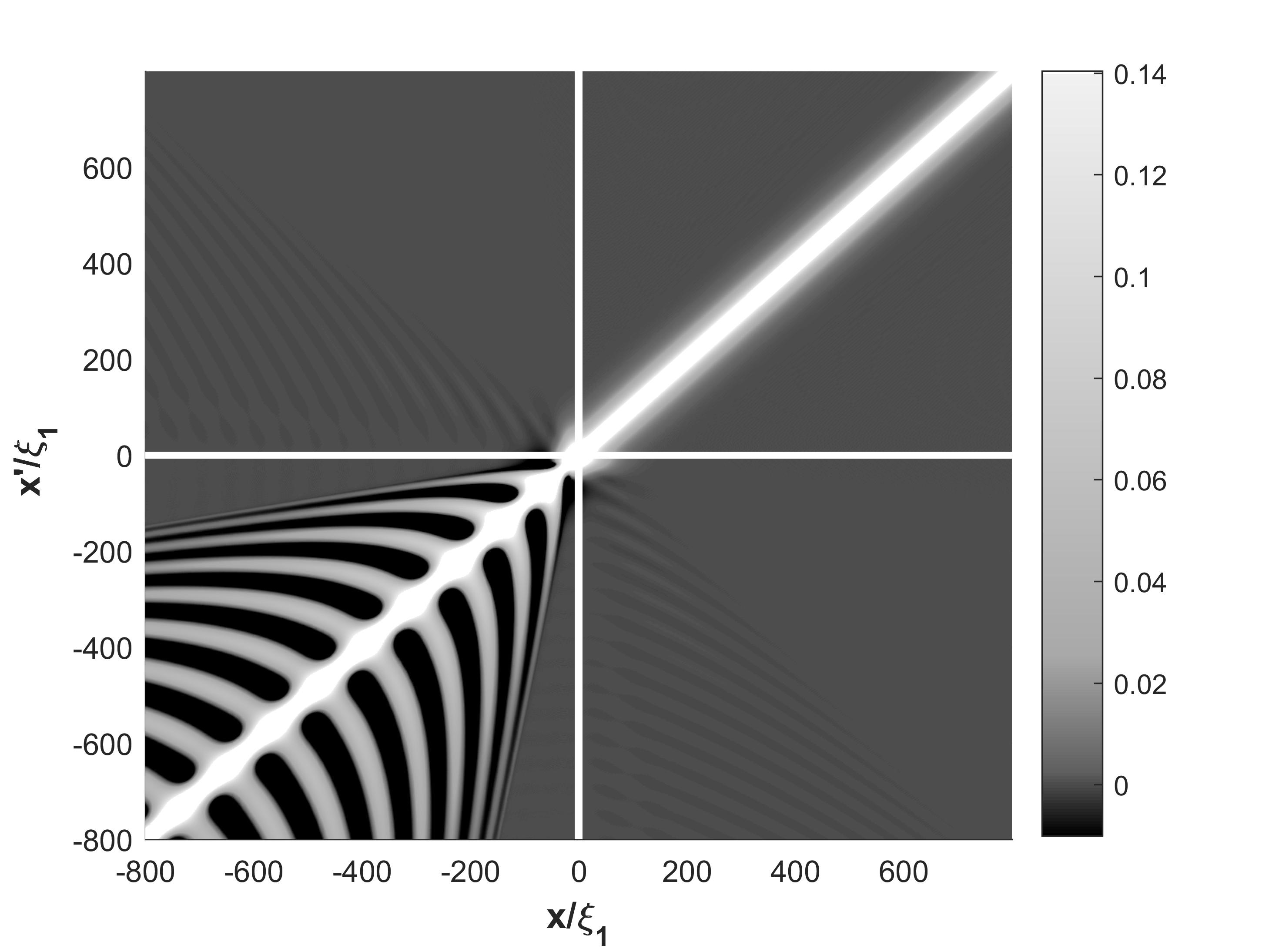}
\includegraphics[width=2.7in]{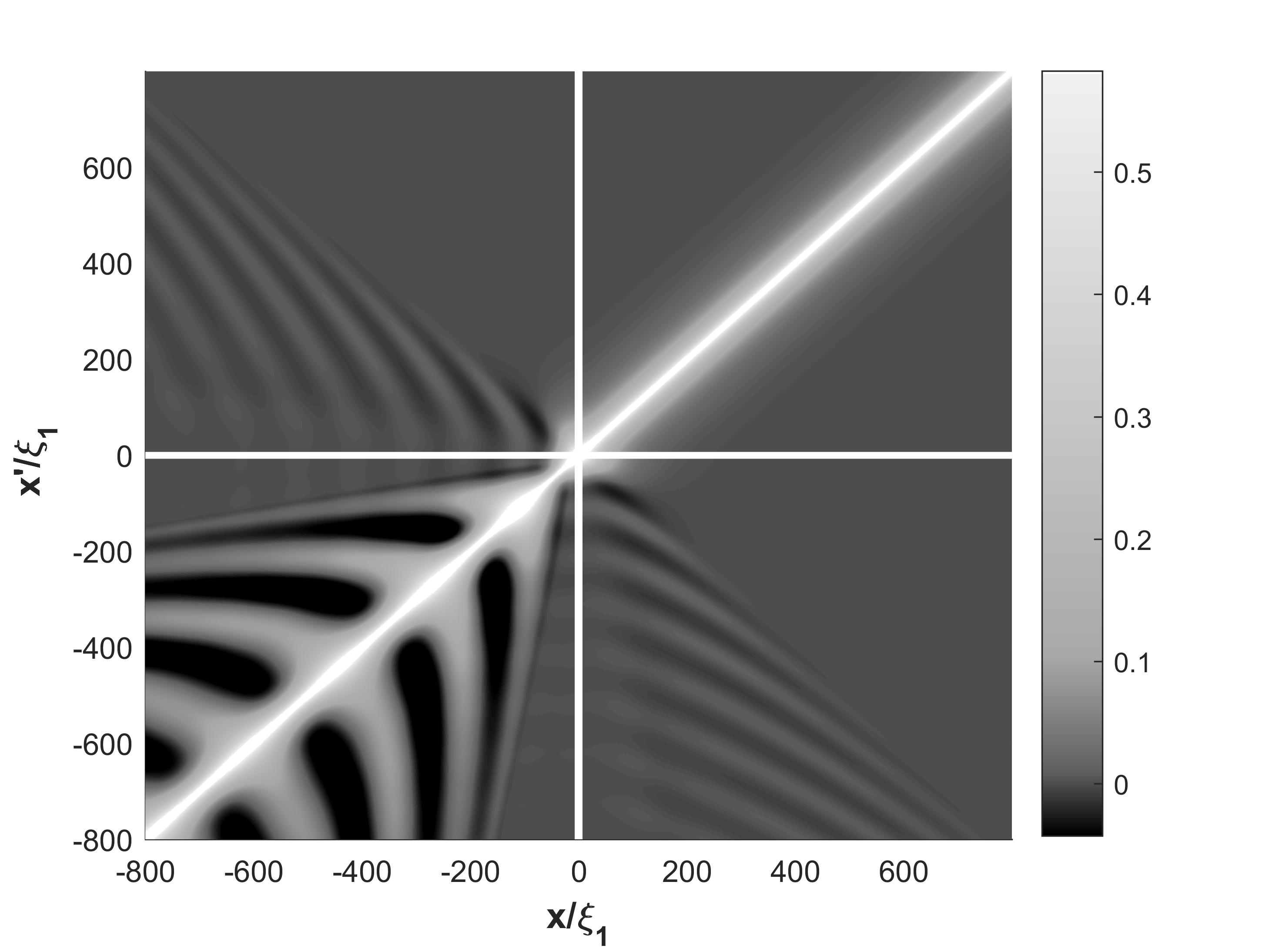}
\includegraphics[width=2.7in]{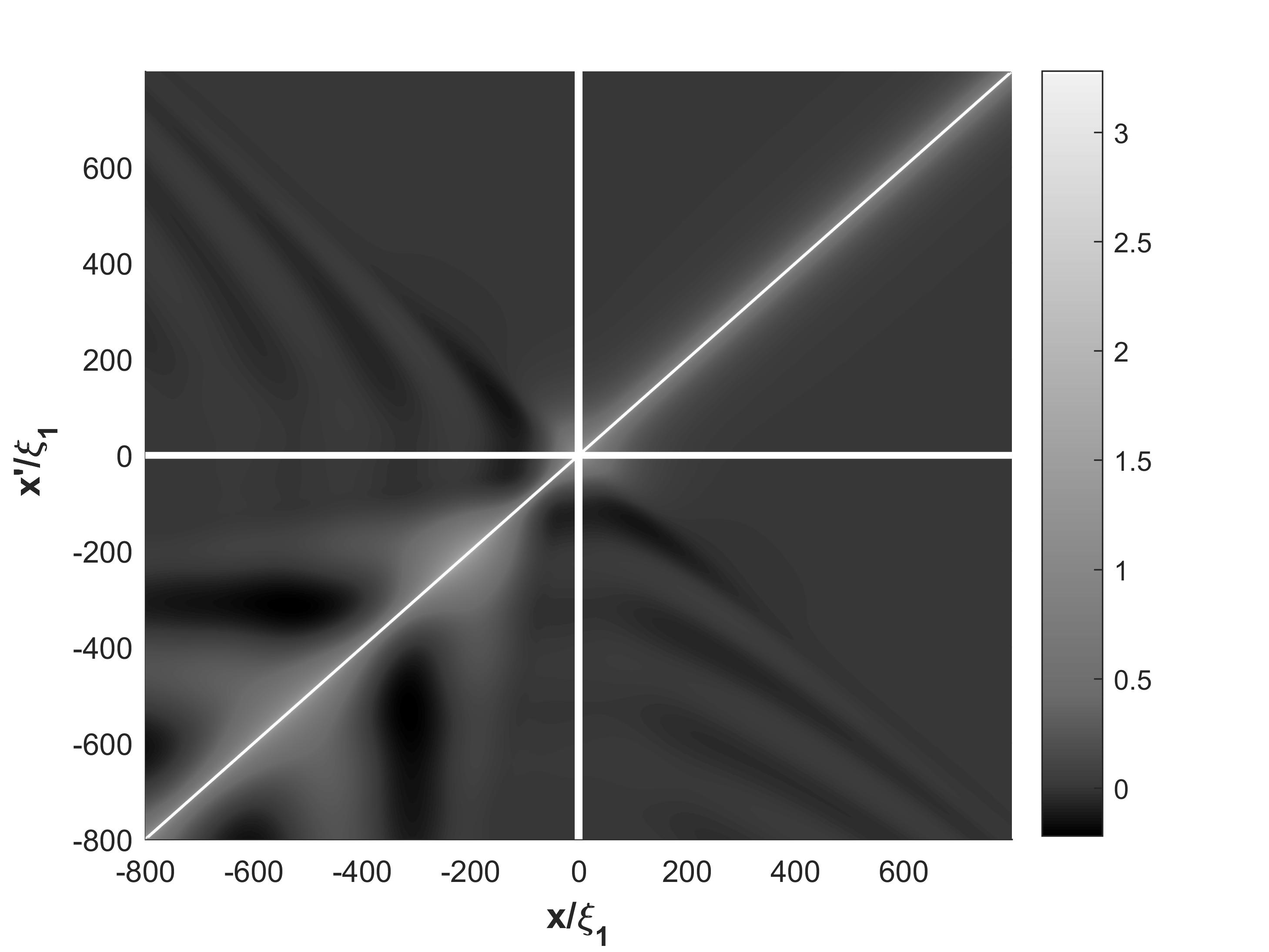}
\includegraphics[width=2.7in]{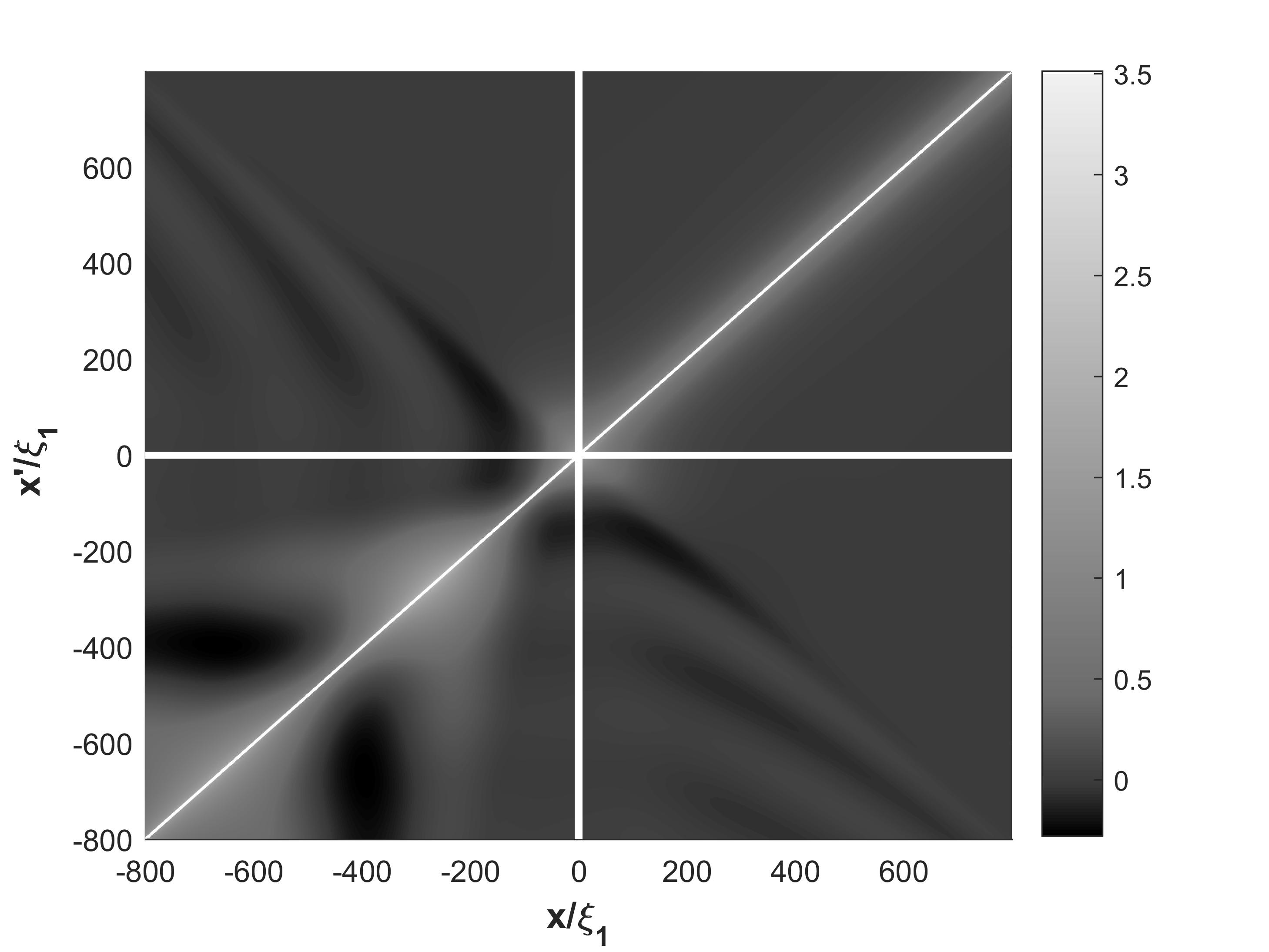}
\includegraphics[width=2.7in]{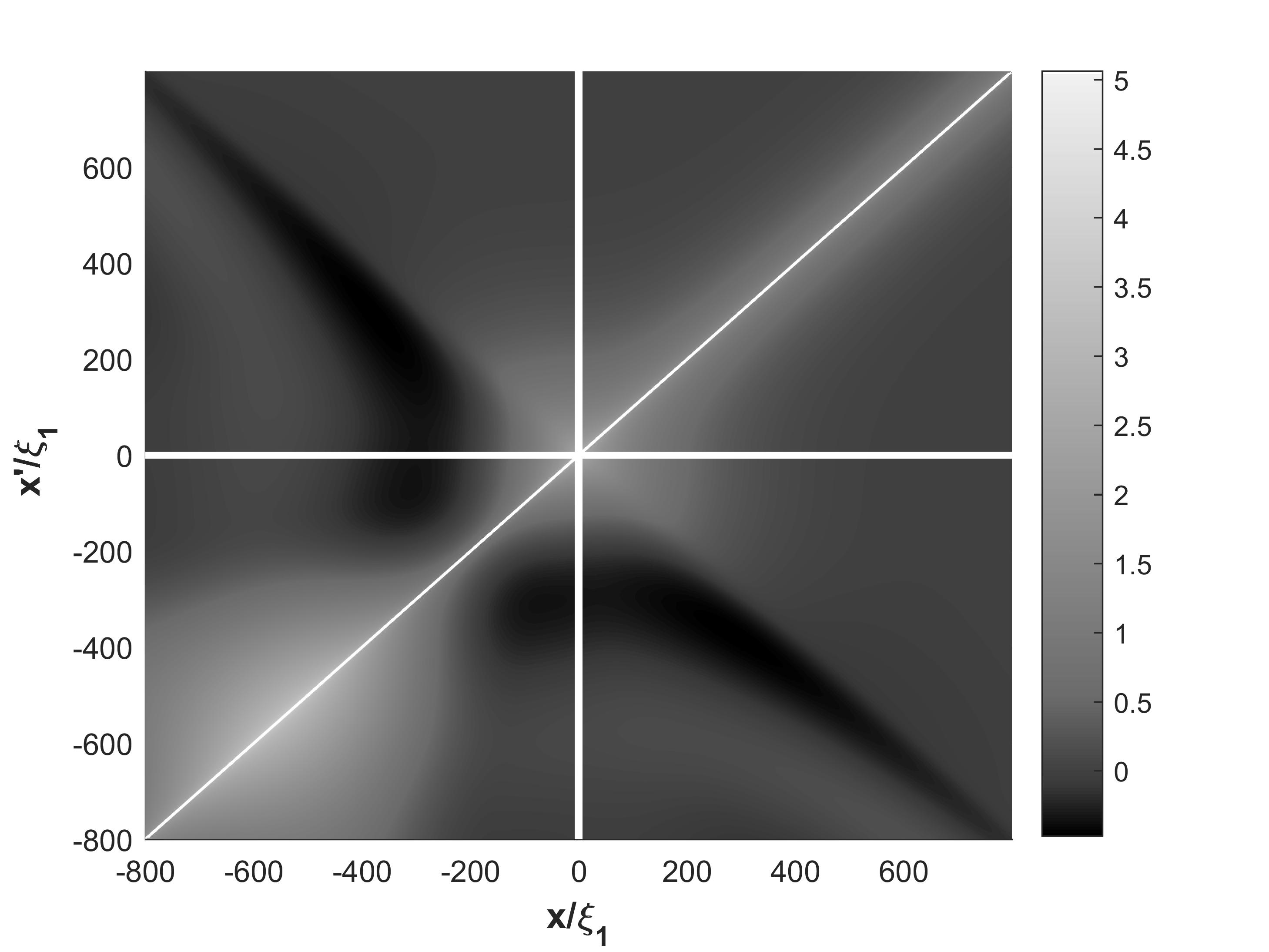}
\includegraphics[width=2.7in]{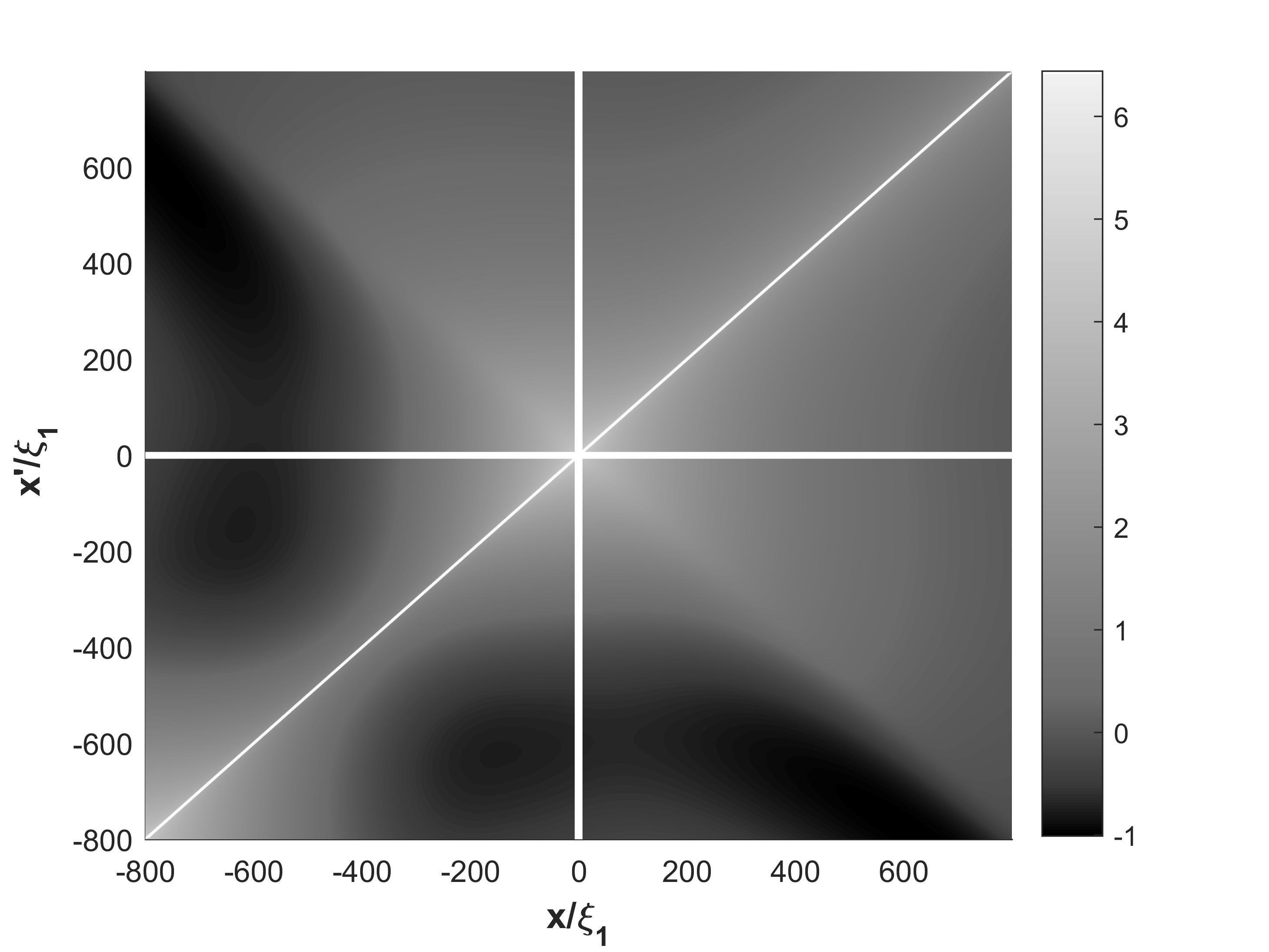}
\includegraphics[width=2.7in]{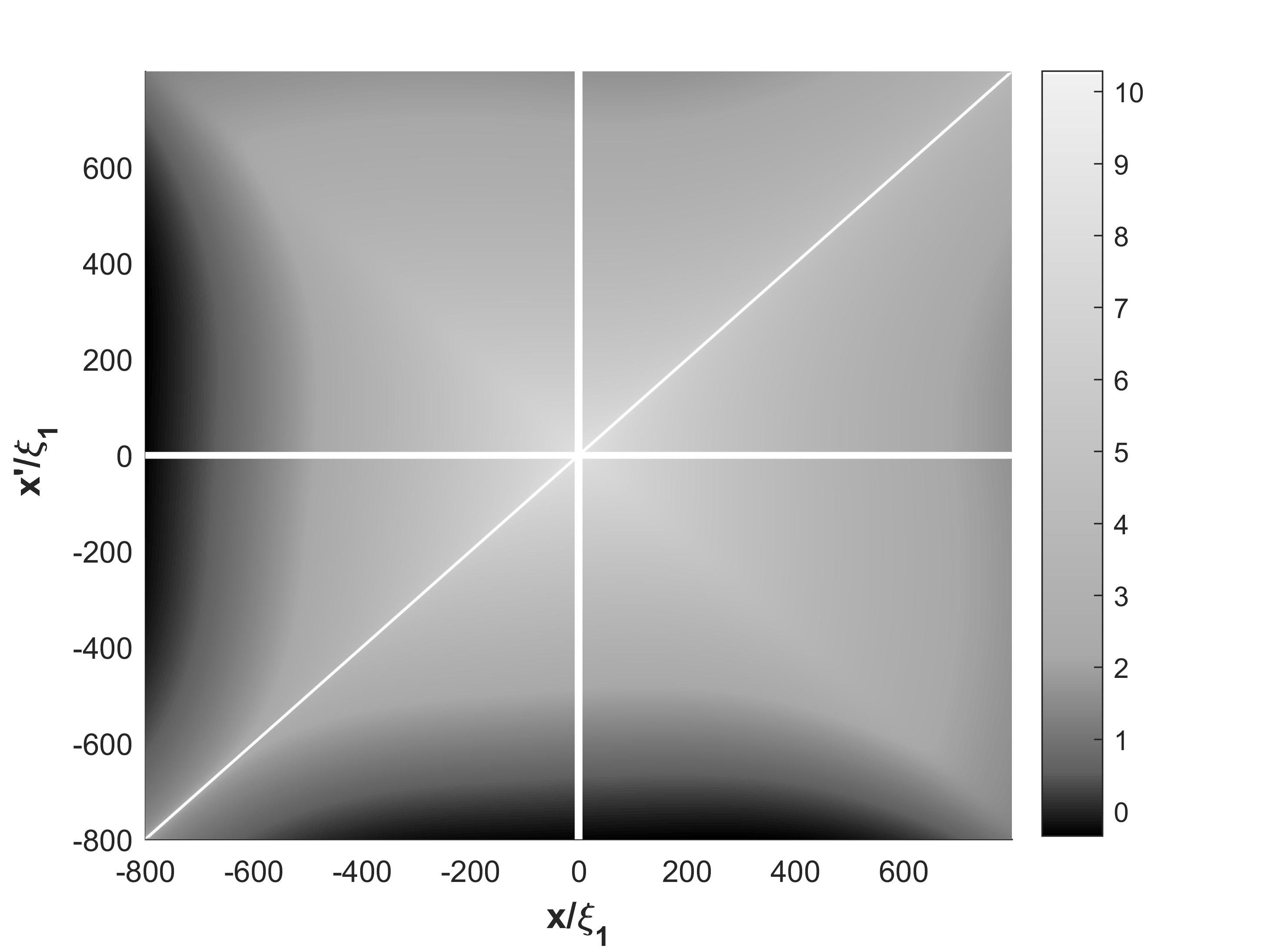}
\includegraphics[width=2.7in]{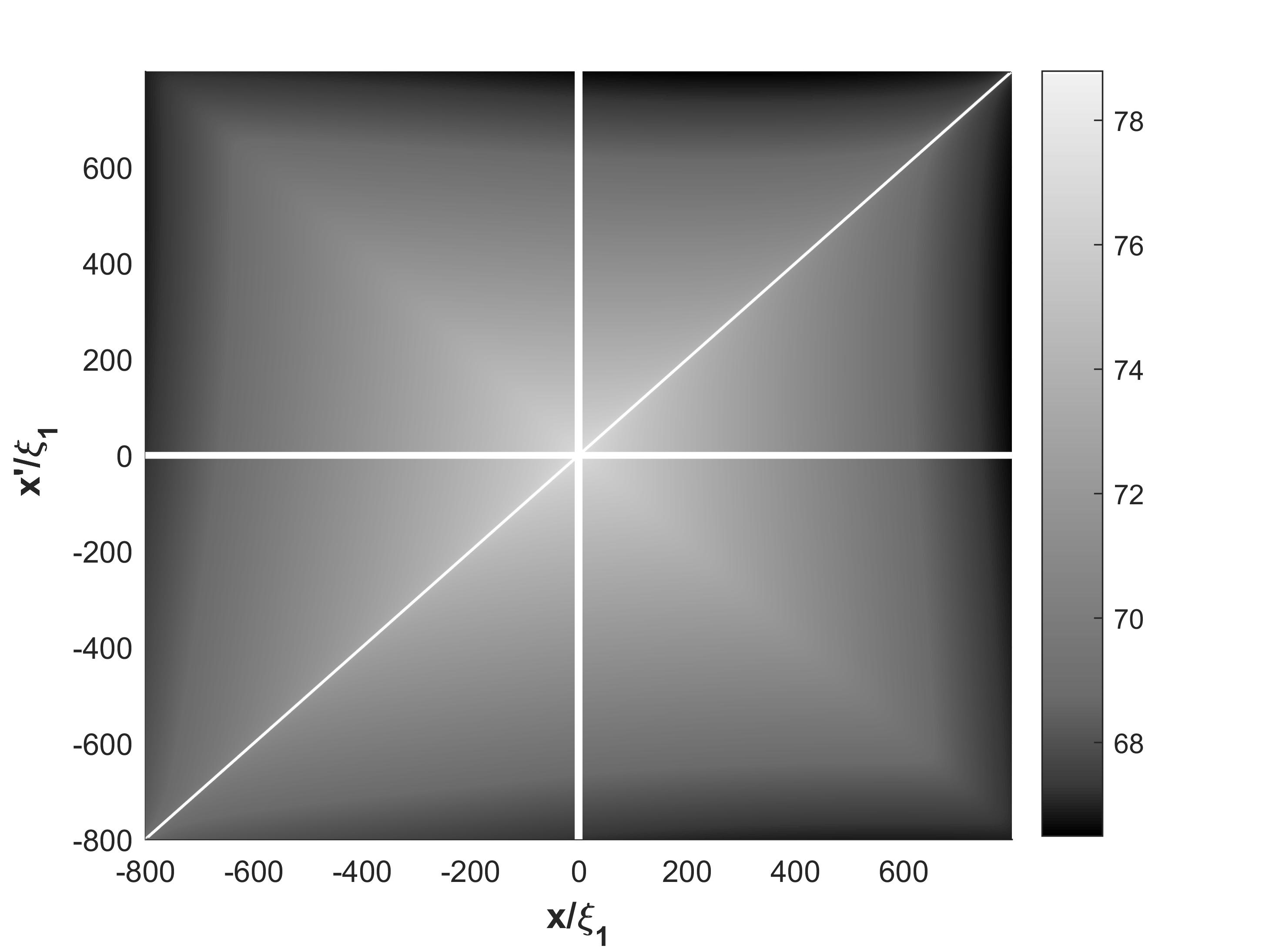}
\caption{\label{Fig:First Plots} Shown are plots for a range of values of the mass $m$ of the phonon field.  From left to right and top to bottom and starting in the upper left the masses are: ${m/m_a=4\times 10^{-2}, 2\times 10^{-2}, 10^{-2}, 8\times 10^{-3}, 6\times 10^{-3}, 4\times 10^{-3}, 2\times 10^{-3}, 10^{-3}}$ and  ${10^{-4}}$. }
\end{figure}

The two-point correlation function for several values of the mass $m$ of the phonons is shown in Fig.~\ref{Fig:First Plots} for a fixed range of values of $x$ and $x'$.

A careful examination of the sequence shows that two things occur.  One is that, for large values of $|x|$ there appears be a scaling of the coordinates of the form $x \to x/m$.
The other is that as $m$ gets small, in the region that is not too far from the horizon, the primary peak appears that occurs in
the massless case when one point is inside and one point is outside the horizon.

\begin{figure}
\includegraphics[width=5.in]{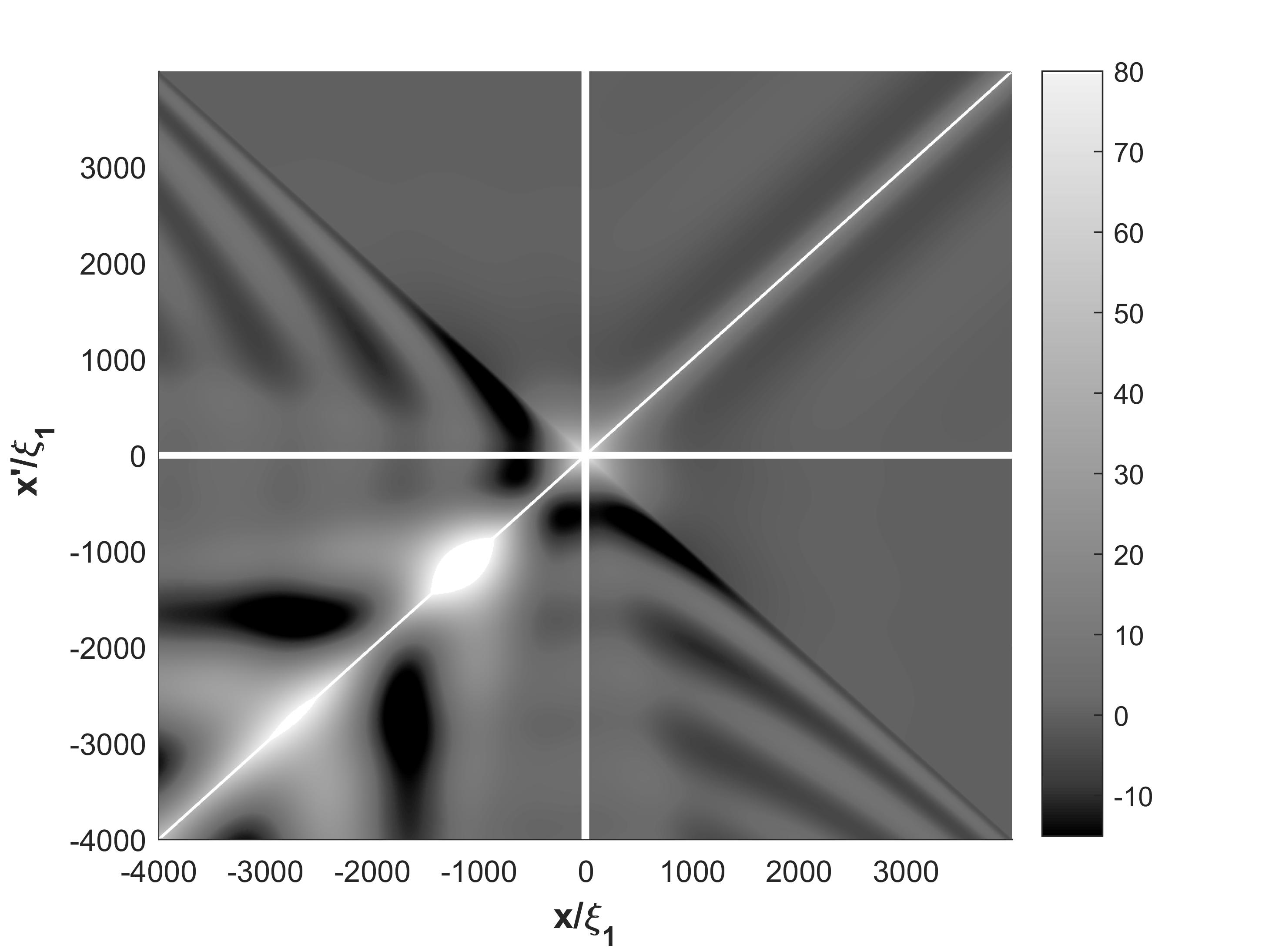}
\caption{\label{Fig:LowMassLargeRange}Plot of the two point function for $m=10^{-4}m_a$, which is the lowest mass case shown in Fig. \ref{Fig:First Plots}(lower right) . The range of this plot has been extended significantly when compared to previous plot to show the structure due to the mass.  }
\end{figure}

The apparent scaling can be seen from the fact that in Fig.~\ref{Fig:First Plots}
the spacing between the peaks increases as $m$ decreases.  As a result the structure moves out of the region shown in the plots.  In Fig.~\ref{Fig:LowMassLargeRange} the two-point correlation function for $m = 10^{-4}m_a$ is shown for a much larger range of values of $x$ and one can see that the structure is still there.
Comparison with the plots in Fig.~\ref{Fig:First Plots} shows that the locations of the peaks appear to scale like $x/m$.

Examination of the mode equation in regions $R$ and $L$,~\eqref{aet1} and~\eqref{aet2} respectively, shows that the mass can be scaled out of these equations if we let $t \to t/m$, $x^* \to x^*/m$, and $\w \to m \w$.  This is reflected in the solutions to the mode equation that are shown in Sec.~\ref{sec:unruh}.  If in those solutions we set $t = \bar{t}/m$, $x^* = \bar{x}^*/m$, and $ \w = m \bar{w}$ then they have the behavior
\be f(\w,m,t,x) = \frac{f(\bar{\w},m=1,\bar{t},\bar{x})}{\sqrt{m}} \label{f-scale}  \;, \ee
where the function $f$ on the left and the function $f$ on the right are the same functions.  Substituting into~\eqref{jj} one finds that the contribution to the two point
function from the integral $J$ is
\be J(m,t-t',x^*,x^{*\;'}) = \frac{J(m=1,\bar{t}-\bar{t'},\bar{x}^*,\bar{x}^{*,'})}{m} \;, \label{J-scale} \ee
where the function on the right and on the left is the same function but with different arguments.
Thus there is an exact scaling of the contribution from $J$ with the mass.

As discussed at the end of Sec.~\ref{sec:unruh}, having used the coordinate system $t$ and $x^*$ related to the metric~\eqref{ams} to compute the two point function we then
change back to the original lab coordinates $T$ and $x$ because for a BEC analogue black hole the density density correlation function (which is computed from the
two point function) is measured experimentally at equal lab times $T$ and $T'$.  It is clear from the relations~\eqref{t-T} and~\eqref{xstar} that so long as
the sound speed varies as a function of position there is no simple linear relationship between $t$ and $x$ or $x^*$ and $x$.  Thus the scaling does not work in
the lab coordinates $T$ and $x$.  However, if both points in the correlation function are far from the horizon then the sound speed is approximately constant.
For example if $x$ is in the $R$ region far from the horizon then
\be c(x) = c_R +\frac{\sigma_v \left(c_L^2-c_R^2\right)}{2
   \pi  c_R x}+O\left(\frac{1}{x}\right)^2 \;. \label{c-exp}\ee
 Multiplying~\eqref{t-T} and~\eqref{xstar} by a factor of $m$ and using~\eqref{c-exp} one finds
\bea \bar{t} &=&  m \left[ T - \frac{v_0}{c_R^2-v_0^2} x - v_0 \frac{\sigma_v (c_R^2-c_L^2)}{\pi (c_R^2-v_0^2)} \log x  + C_1 + O\left(\frac{1}{x}\right) \right] \nonumber \\
     \bar{x}^* &=& m \left[  \frac{c_R}{c_R^2-v_0^2} x + \frac{\sigma_v (c_R^2-c_L^2)(c_R^2+v_0^2)}{2 \pi c_R (c_R^2-v_0^2)^2} \log x + C_2 + O\left(\frac{1}{x}\right) \right]\;, \eea
with $C_1$ and $C_2$ constants.
One finds similar expressions if $x$ is in the $L$ region.  Clearly it is not possible to eliminate the $m$ dependence of the right-hand sides through a simple scaling
of $x$.  However, if one changes to $x = \bar{x}/m$ and $T = \bar{T}/m$, then the leading order terms on the right-hand sides of these equations do not depend on $m$.  Further all of the
other terms decrease as $m$ does for fixed $\bar{x}$. If these expansions are substituted into the expression for $J(m=1,\bar{t},\bar{x}^*)$ then it is found that the
dependence on the mass becomes weaker for fixed $\bar{x}$ as the mass decreases.  So we have an approximate scaling of this contribution to the two point function for the
point separations that we are interested in.

If~\eqref{f-scale} is substituted into~\eqref{zuz} one sees that there is no simple scaling for the contribution to the two point function $I$.
The reason is that we are not changing the sound speed profile~\eqref{c-used} as we change the mass and therefore the surface gravity $\kappa$ does not change.  However the locations of the correlation peaks should depend on the arguments of the exponentials of the mode functions and not very much on the functions ${\rm csch}(\pi \w/\kappa)$ and $\coth(\pi \w/\kappa)$ which do not oscillate.  Therefore we would expect that the locations of the peaks would scale with $x$ in the same approximate way for $I$ as they do for $J$.

\begin{figure}
\includegraphics[width=4.in]{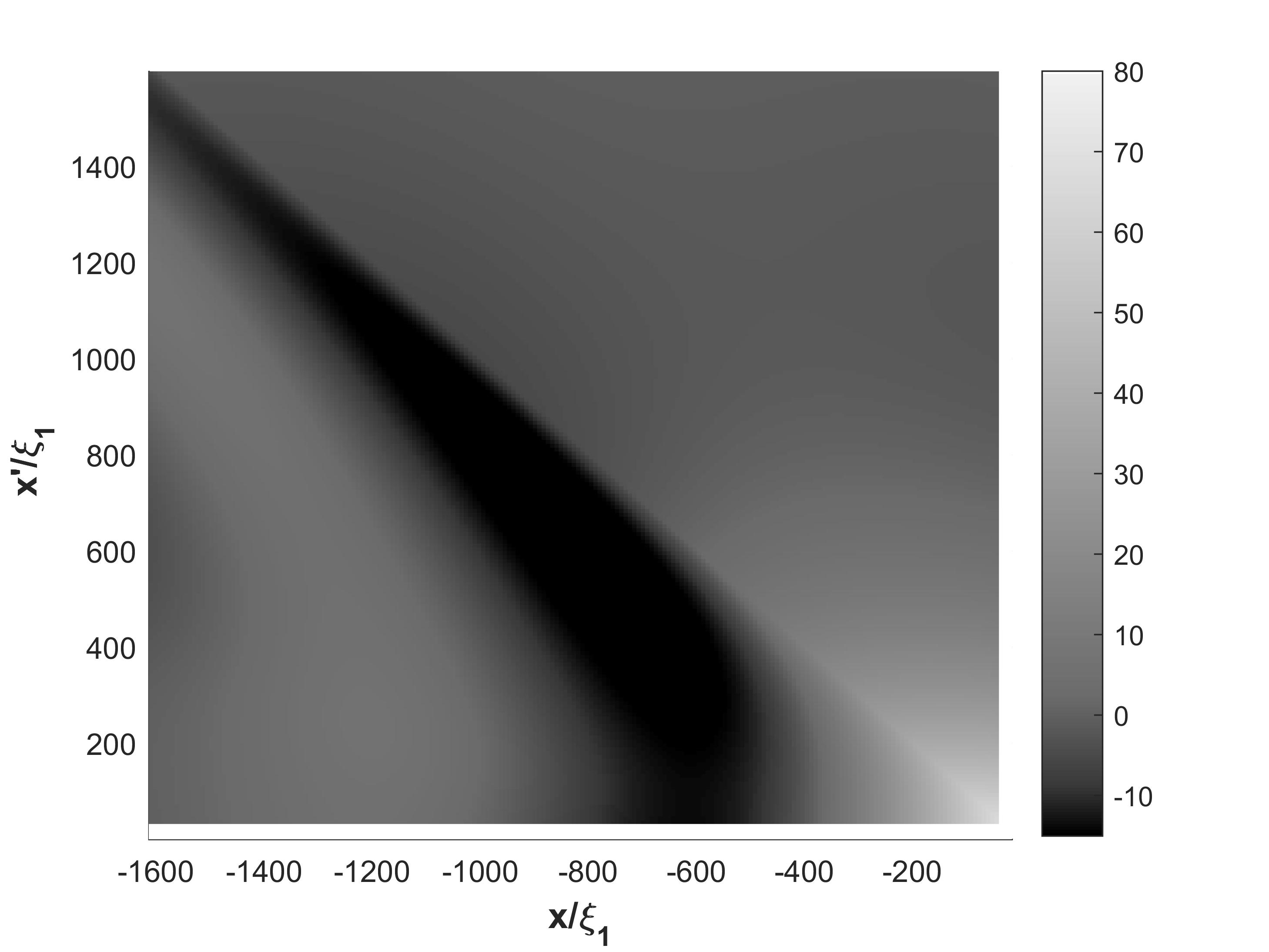}
\caption{\label{Fig:LowMassHPPeak}Plot of the two point function for $m=10^{-4}m_a$, which is the lowest mass case shown in Fig.~\ref{Fig:First Plots}(lower right).  It is
also the same mass as was used for the plot in Fig.~\ref{Fig:LowMassLargeRange}. The region shown is where one point is in the interior and one point is in the exterior of the analogue BH. Near the horizon a peak can be seen at the same approximate location as peak 1) in the massless case, see Eq.~\eqref{m0-1} and Fig.~\ref{Fig:2013PRD BEC DD Plot}.    This peak disappears for points which are further from the horizon.      }
\end{figure}

In the limit $m \to 0$, the primary correlation peak with one point inside and one point outside the horizon is the only one that should occur in our model
since the effective potential is zero.  In fact in Fig.~\ref{Fig:First Plots} it appears that in the finite sized region shown, this peak appears once the mass is small enough.
However, as seen in Fig.~\ref{Fig:LowMassHPPeak} this peak does not remain a peak when both points are far
enough away from the horizon.  Thus in some sense the two point function for very small masses is qualitatively similar
to the two point function in the massless case only if one's attention is restricted to a finite region centered around the horizon.

To understand why this peak appears in a finite region containing the horizon for small $m$ it is sufficient to consider the primary contribution to the peak which comes from the $f^H_R$ and the $f^H_L$ modes.
Recall that in our calculations the scattering occurs at $x^* = 0$ in $R$ and the particle creation occurs at $x^* = 0$ in $L$.  For $x^* \ge 0$
$f^H_R$ is given by~\eqref{mnmn} and $f^L_R$ is given by~\eqref{lulu}.
The part of $f^H_L$ that contributes to the primary in-out peak if $m = 0$ is the part proportional to $e^{-i k_L x^{* \; '}} $.  The contribution of both of these
modes to the integral $I$ in~\eqref{zuz} for  $x^* > 0$ and $x^{*\;'} > 0$ is
\be I_1 = \frac{1}{2 \pi} \int_{m_R}^\infty \frac{d \w}{\sinh \left(\frac{\pi \w}{\kappa} \right)} \frac{(k_L+\w)}{k_L (k_R+\w)}  \cos[\w (t-t') + k_R x^* - k_L x^{* \; '}]  \;. \label{I1-2} \ee
In the massless case $k_L = k_R = \w$ and the correlation peak occurs at points where the argument of the cosine vanishes.\footnote{In the massless case one must impose an infrared cutoff on the integral in~\eqref{I1-2}.  The cutoff
will impose structure in the correlation function for large enough
separations of the points, but if the density density correlation function
is computed then the cutoff can be taken to zero and the extraneous
structure disappears.}  These are given by the condition $t-t' = -(x^* - x^{*\;'})$.
In the low mass limit
\bea k_R &=& \w - \frac{m_R^2}{2 \w} + \ldots   \nonumber \\
     k_L &=&  \w +  \frac{m_L^2}{2 \w} + \ldots \;.  \eea
Thus for modes with $m^2 \ll \w$ the argument of the cosine in~\eqref{I1-2} does not vary rapidly with $\w$ near the position of the peak in the massless case for
values of $x^*$ and $x^{*\;'}$ such that $\frac{m_R^2}{2 \w} x^* \ll 1$ and $\frac{m_L^2}{2 \w} x^{*\;'} \ll 1$.   Thus we would expect the peak to occur
for values of $x^*$ and $x^{*\;'}$ which satisfy these conditions.  However for significantly larger values of $x^*$ and $x^{*\;'}$ these terms in
the argument will vary significantly with $\w$ and this change in the phase of the cosine in~\eqref{I1-2} with respect to $\w$ will result in cancellations that eliminate
this peak when the integral is computed.
This is exactly what is
observed for large distances from the horizon in Fig.~\ref{Fig:LowMassHPPeak}.

\section{Conclusions}
\label{sec:conclusions}

For an acoustic black hole we have presented a pedagogical toy model that shows the catalytic effect which the presence of excitations transverse to the flow has in producing a rather rich and complex structure in the correlation functions, in particular in the interior (BH) region. Transverse excitations induce a masslike term in the equation for the longitudinal modes which not only causes backscattering of the modes but also allows the existence of zero frequency standing waves (undulations) inside the horizon.

We have found a few different groups of correlation peaks that lie along approximately parallel curves.  In contrast, there are only three prominent correlation peaks in the
density density correlation function in the massless case.  In the massive case their locations are not peaks for the most part.  However when both points are inside
the horizon the locations of the correlation peaks in the massless case serve as boundaries for a group of prominent peaks in the massive one.
When one point is inside and one point is outside the horizon, for the calculations we have done, there are boundaries at the approximate, but not exact, locations
of the correlation peaks in the massless case.

The most prominent group of peaks in the massive case occur when both points are inside the horizon.  They have approximately the same distances between them as the undulations do, but they have a different pattern.  This means  that the contribution to the two point function from the undulations is effectively erased by the contributions of modes with higher frequencies.  The boundary of this prominent group of peaks on both sides corresponds to the locations of the peak that occurs in the massless case when both points are inside the horizon.

If one point is inside and one point is outside the horizon one group of peaks is close to the horizon and roughly parallel to it.  Their upper boundary
 is, for the calculations we did, in close proximity with the location of the secondary correlation peak that occurs in the massless case.   A
second group of peaks which are roughly diagonal has an upper boundary which is in close proximity with the location of the primary correlation peak in the massless case.

There are also very small correlation peaks which do not show up in our plots.  These come from terms in the two point function
for which there is no stationary phase approximation for the given integration range.  They occur in each quadrant of Fig.~\ref{Fig:ExtremeRange}.  Because they are so small it would be extremely difficult to observe them experimentally
and it is likely that other effects which are not accounted for in our simple model would be more important.

We have provided strong evidence that there is a discontinuity between the types of structure that occur in the massless and massive cases if one considers
the behavior of the two-point correlation function over the entire analogue spacetime.  However there appears to be a smooth transition to the massless case if attention is
restricted to a finite sized region.   There are two reasons for this.  First we have provided evidence that all of the structures observed scale like $1/m$ so that the peaks get father apart as the mass decreases.    Second we have also shown that
a single correlation peak appears near the horizon when the mass gets small enough and one point is inside one is outside the horizon.  At large distances from the horizon this peak disappears.  However, if attention is restricted to a finite region that contains the origin, then for small enough masses this peak is qualitatively similar to the primary correlation peak found in the massless case.

On physical grounds we expect the correlation peaks that appear in our plots to be quite general and not to be an artifact of the simple model used.
For BEC analogue black holes, experimental evidence has been reported  only for the main peak in the massless case which occurs when one point is inside and one point is outside the horizon. However in future experiments, where the sensitivity in the correlation function measurements will be increased, one expects the various patterns we have discussed to show up in systems with excitations of one or more transverse modes, putting the QFT in curved space approach to this peculiar type of condensed matter system on even more solid ground.

\acknowledgments
We would like to thank Jeff Steinhauer and Germain Rousseaux for interesting and useful comments.
This work is partially supported
by the Spanish Ministerio de Econom\'ia, Industria y Competividad Grant numbers FIS2014-57387-C3-1-P, FIS2017-84440-C2-1-P, the Generalitat
Valenciana project SEJI/2017/042, the Severo Ochoa Excellence Center Project
SEV-2014-0398, the National Science Foundation under Grants
No. PHY-0856050, No. PHY-1308325, and No. PHY-1505875 to Wake Forest University, and a Wake Forest University Bridge Grant.
Some of the numerical work was done using the WFU DEAC cluster; we thank the WFU Provost's Office and Information Systems Department
for their generous support.

\clearpage
\begin{appendix}

\section{Connection with a BEC analogue black hole}
\label{appendix-connection}

In~\cite{paper2013} a detailed derivation is given for the equation satisfied by the phase fluctuation operator $\hat{\theta}_1$ which along with the density fluctuation
operator $\hat{n}$ describes the noncondensed part of the BEC.  In the hydrodynamical approximation (see~\cite{paper2013}) the equation is
\be \Box \hat{\theta}_1 = 0  \;, \ee
with the d'Alembertian evaluated using the four dimensional metric
\be ds^2 = \frac{n}{m_a c} \left[ - c^2 dT^2 + (dx + v_0 dT)^2 + dy^2 + dz^2 \right]  \;. \label{metric-4D} \ee
Here $n$ is the density of the condensate (assumed to be constant), $m_a$ is the mass of one of the atoms in the condensate, and we have written the form of the metric for the case we are considering which is $\vec{v} = - v_0 \hat{x}$.
With the change of variable~\eqref{t-T}  the metric and the wave equation are
\bes \bea ds^2 &=& \frac{n}{m_a c} \left[-(c^2-v_0^2) d t^2 + \frac{c^2}{c^2-v_0^2} dx^2 + dy^2 + dz^2 \right]  \;, \\
          \Box \hat{\theta}_1  &=& \left[ -\frac{c}{c^2-v_0^2} \partial_t^2 + c \partial_x \left(\frac{c^2-v_0^2}{c^2} \partial_x \right) + c (\partial_y^2 + \partial_z^2) \right] \hat{\theta}_1 = 0 \;.   \label{theta1-eq}  \eea \ees

Next we perform a dimensional reduction by defining a new field $\hat{\theta}^{(2)}_1(x,t)$ such that
 \be \hat{\theta}_1 = \sqrt{\frac{m_a c}{n \ell_\perp^2}} \; \hat{\theta}^{(2)}_1(x,t) \exp \left[i (k^\perp_y y + k^\perp_z z)\right]   \;, \ee
 with $\ell_\perp$ the transverse dimension of the BEC.  Substituting into~\eqref{theta1-eq} gives
\be  \left[ - \frac{c^{3/2}}{c^2-v_0^2} \partial_t^2 + c \partial_x \left( \frac{(c^2-v_0^2)}{c^2} \partial_x \sqrt{c}\right) -c^{3/2} (k^\perp)^2 \right] \hat{\theta}^{(2)}_1   = 0  \;, \ee
with 
\be
(k^\perp)^2=(k^\perp_y)^2+(k^\perp_z)^2.\;
\ee
This can be written in the form
  \be [\Box - c \, m^2 + V(x)] \hat{\theta}^{(2)}_1 = 0  \;, \ee
  with
 \be \Box = - \frac{c}{c^2-v_0^2} \partial_t^2 + \frac{c^2-v_0^2}{c} \partial_x^2 + \frac{d c}{dx} \left(1+ \frac{v_0^2}{c^2} \right) \partial_x \;, \ee
  the two dimensional d'Alembertian for the metric
  \be ds^2 = -\frac{c^2-v_0^2}{c} dt^2 + \frac{c}{c^2-v_0^2} dx^2  \;, \ee
  and
  \bea m^2  &=&  (k^\perp)^2  \nonumber \\
       V(x) &=& \frac{1}{2} \frac{d^2 c}{dx^2} \left(1  - \frac{v_0^2}{c^2} \right) -\frac{1}{4 c} \left(\frac{d c}{dx}\right)^2 +  \frac{5 v_0^2}{4 c^3} \left(\frac{dc}{dx} \right)^2 \;. \eea
  The final form of the metric and the wave equation is obtained by using~\eqref{xstar} to obtain
\bes  \bea & & \left[- \partial_t^2 + \partial_{x^*}^2  - m^2 (c^2 - v_0^2)  + \frac{(c^2-v_0^2)}{c} V(x) \right] \hat{\theta}^{(2)}_1 = 0  \;. \label{mode-eq-2D-2013} \\
  & & ds^2 =  \frac{c^2-v_0^2}{c} [ -\partial_t^2 + \partial_{x^*}^2 ]  \;,  \label{metric-2D-2013} \eea  \ees

The two dimensional metrics~\eqref{metric-2D-2013} and~\eqref{ams} differ by an overall factor of $c^{-1}$.   However they both have horizons at the location where $c = v_0$
and they both have the same surface gravity,~\eqref{kappa}, if the horizon is at $x = 0$.  Further
if we set $V = 0$ then the equation satisfied by $\hat{\theta}^{(2)}_1$~\eqref{mode-eq-2D-2013} is the same as that satisfied by $\hat{\phi}$~\eqref{kgt}.
Since the relationship between $x$ and $x^*$ is also the same, this means that the equations for the two point function
in the Unruh state~\eqref{ii},~\eqref{jj}, and~\eqref{zuz} are the same for both fields if the same sound speed profile is used.  This is what allows us to use our simple
model for a massive scalar field propagating in a two dimensional spacetime to obtain information that is relevant for a BEC analogue BH when
there are transverse excitations of the phonon modes.

\section{${f_I}$ mode normalization \label{SCRIminusNomalization}}

The normalization for the modes $f_I$ is nontrivial because the masslike term in the mode equation varies with position.  For our toy model, in both regions $R$ and $L$ this term is zero for $x^* < 0$ and a nonzero constant for $x^* >0$.

The $f_I$ modes can be written in the form
\be f_I =  N e^{- i \w t} \chi_I(x)  \;, \label{fI-norm} \ee
with $N$ a normalization constant.  $\chi_I$ satisfies the same equation~\eqref{chi-eq} as $\chi^H_R$.
In $R$ the modes can be thought of as $v$ modes which originate at $x = +\infty$ and initially move inward.
In terms of $\chi_I$, a massive incoming $v$ mode, $e^{-ik_Rx^*}$, is partially reflected back to $x=+\infty$ by the boundary at $x^*=0$
in the form $B_\w e^{ik_R x^*}$ ($u$ mode) and is partially transmitted as a massless mode towards the horizon in the form
$C_\w e^{-i\omega x^*}$ with $k_R$ defined in~\eqref{kR-def}.  Continuity of $\chi_I$ and its first derivative at $x^*=0$ then gives
\bes \bea B_\w &=& \frac{k_R-\omega}{k_R+\omega}  \ , \label{eq:SCRICoeff1} \\
C_\w & = & \frac{2\sqrt{\omega k_R}}{k_R+\omega}\ .\label{eq:SCRICoeff2} \eea \label{BC}  \ees

  We normalize by choosing the $t = 0$ surface in region $R$ along with the part of the past horizon $H^{-}$ which is in region $L$ as the Cauchy surface along which we evaluate the scalar product~\eqref{sc}.  Note that the $f_I$ modes vanish on $H^-$.
 Then
 \bes \bea
 \left(f_{{I}}(\omega, x), f_{{I}}(\omega^\prime, x)\right)&=&-i\int_{-\infty}^{\infty} f_{{I}}(\omega, x^*)\overset{\leftrightarrow}{\partial_t} f_{{I}}^{*}(\omega^\prime, x^*)dx^*
 \label{fI-4} \\
 \nonumber \\
 &=&  (\omega+\omega^\prime)\left[\int_{-\infty}^0 C_{\omega}N_{\omega} C_{\omega^\prime}^*N_{\omega^\prime}^*e^{-i x^*(\omega-\omega^\prime)}dx^*\right. \nonumber \\
&&    \quad+ \int_0^{\infty}\left( B_{\omega}N_{\omega} B_{\omega^\prime}^*N_{\omega^\prime}^*e^{i x^*(k_R -k_R^\prime)}+ N_{\omega} N_{\omega^\prime}^*e^{-i x^*(k_R -k_R^\prime)}\right.\nonumber \\
&& \quad \left.\left.+ B_{\omega}N_{\omega}  N_{\omega^\prime}^*e^{i x^*(k_R +k_R^\prime)}+ N_{\omega} B_{\omega^\prime}^* N_{\omega^\prime}^* e^{-i x^*(k_R +k_R^\prime)}  \right)dx^*\right]. \label{fI-scalar}  \eea  \ees

The first integral in~\eqref{fI-scalar} can be computed using an integrating factor of the form $e^{\epsilon x^{*}}$, with $0 < \epsilon \ll 1$, along with the relation~\cite{matwal}
\be \frac{1}{x - x_0 \mp i \epsilon} = P \frac{1}{x-x_0}  \pm i \pi \delta(x-x_0) \;, \ee
with $P$ denoting the principle value.  Dropping the ``P'' we find
 \bea
& & \int_{-\infty}^{x_{0}^*} C_{\omega}N_{\omega} C_{\omega^\prime}^*N_{\omega^\prime}^*e^{-i x^*(\omega-\omega^\prime)}dx^*=\lim_{\epsilon \to 0}\int_{-\infty}^{x_{0}^*}C_{\omega}N_{\omega} C_{\omega^\prime}^*N_{\omega^\prime}^*e^{-i x^*(\omega-\omega^\prime+i\epsilon)}dx^* \nonumber \\
& & \;\;\;\;\; =\lim_{\epsilon \to 0}\left[ C_{\omega}N_{\omega} C_{\omega^\prime}^*N_{\omega^\prime}^*\left(\frac{1}{-i(\omega-\omega^\prime+i\epsilon)}\right)\right]
\nonumber \\
& &  \;\;\;\;\; =\left[C_{\omega}N_{\omega} C_{\omega^\prime}^*N_{\omega^\prime}^*\left(\frac{i}{(\omega-\omega^\prime)}+\pi \delta(\omega-\omega^\prime)\right)\right] \;.
 \eea
 The second integral can be computed in a similar manner.

Using the identity ${\delta(k_R-k_R^\prime)=\frac{k_R}{\omega}[\delta(\omega-\omega^\prime)-\delta(\omega+\omega^\prime)]}$ along with the fact that ${\delta(\omega+\omega^\prime)=0=\delta(k_R+k_R^\prime)}$ it is easy to show that
 \bea
 \left(f_{{I}}(\omega, x), f_{{I}}(\omega^\prime, x)\right)&=& (\omega+\omega^\prime)\left[C_{{\omega}} N_{\omega} C_{\omega^\prime}^* N_{\omega^\prime}^*\left(\frac{i}{(\omega-\omega^\prime)}+\pi \delta(\omega-\omega^\prime)\right) \right.  \nonumber \\
  & &  \left.  + B_{\omega} N_{\omega} B_{\omega^\prime}^*  N_{\omega^\prime}^* \left(\frac{i}{(\omega-\omega^\prime)}+\pi \frac{k_R}{\omega} \delta(\omega-\omega^\prime)\right) \right. \nonumber \\
 & & \left. +  N_{\omega} N_{\omega^\prime}^*\left(\frac{-i}{(\omega-\omega^\prime)}+\pi\frac{k_R}{\omega} \delta(\omega-\omega^\prime)\right) +B_{\omega}  N_{\omega} N_{\omega^\prime}^* \left(\frac{i}{(\omega-\omega^\prime)}\right) \right. \nonumber \\
   & & \left. + N_{\omega}B_{\omega^\prime}^* N_{\omega^\prime}^*\left(\frac{-i}{(\omega-\omega^\prime)}\right)     \right].
 \eea
 Using Eqs.~\eqref{BC} for $B_\w$ and $C_\w$ gives
  \bea & & \left(f_{{I}}(\omega, x), f_{{I}}(\omega^\prime, x)\right)=
   (\omega+\omega^\prime) N_{\omega} N_{\omega^\prime}^*\left\{\left(\frac{2}{1+\frac{\omega}{k_R}}\right)\left(\frac{2}{1+\frac{\omega^\prime}{k_R^\prime}}\right)\left[\frac{i}{(\omega-\omega^\prime)}+\pi \delta(\omega-\omega^\prime)\right]  \right. \nonumber \\
 \;\;\;\;\; & &  \left. + \left(\frac{1-\frac{\omega}{k_R}}{1+\frac{\omega}{k_R}}\right) \left(\frac{1-\frac{\omega^\prime}{k_R^\prime}}{1+\frac{\omega^\prime}{k_R^\prime}}\right) \left[\frac{i}{(k_R-k_R^\prime)}+\pi \frac{k_R}{\omega} \delta(\omega-\omega^\prime)\right]+  \left[\frac{-i}{(k_R-k_R^\prime)}+\pi\frac{k_R}{\omega} \delta(\omega-\omega^\prime)\right] \right. \nonumber \\
 & & \;\;\;\;\; \left. + \left(\frac{1-\frac{\omega}{k_R}}{1+\frac{\omega}{k_R}}\right) \left[\frac{i}{(k_R-k_R^\prime)}\right]+ \left(\frac{1-\frac{\omega^\prime}{k_R^\prime}}{1+\frac{\omega^\prime}{k_R^\prime}}\right)\left[\frac{-i}{(k_R-k_R^\prime)}\right]     \right\}\;. \label{eq:NormalizationEquation}
 \eea
The right-hand side of this equation can be broken into a part which is a factor of ${\delta(\omega-\omega^\prime)}$ and a part which is not. The latter is
 \bea   & & (\omega+\omega^\prime) N_{\omega} N_{\omega^\prime}^*\left\{\left(\frac{2}{1+\frac{\omega}{k_R}}\right)\left(\frac{2}{1+\frac{\omega^\prime}{k_R^\prime}}\right)\left[\frac{i}{(\omega-\omega^\prime)}\right]+
  \left(\frac{1-\frac{\omega}{k_R}}{1+\frac{\omega}{k_R}}\right) \left(\frac{1-\frac{\omega^\prime}{k_R^\prime}}{1+\frac{\omega^\prime}{k_R^\prime}}\right) \left[\frac{i}{(k_R-k_R^\prime)}\right]  \right. \nonumber \\
  & & \left. +  \left[\frac{-i}{(k_R-k_R^\prime)}\right] + \left(\frac{1-\frac{\omega}{k_R}}{1+\frac{\omega}{k_R}}\right) \left[\frac{i}{(k_R-k_R^\prime)}\right]+ \left(\frac{1-\frac{\omega^\prime}{k_R^\prime}}{1+\frac{\omega^\prime}{k_R^\prime}}\right)\left[\frac{-i}{(k_R-k_R^\prime)}\right]     \right\}    \;. \label{eq:goes to zero}
 \eea
 After some algebra one finds that this expression is equal to zero.  The part of the Eq. \ref{eq:NormalizationEquation} which is proportional to  ${\delta(\omega-\omega^\prime)}$ is,
   \bea
    & & \pi (\omega+\omega^\prime) N_{\omega} N_{\omega^\prime}^*\left\{\left(\frac{2}{1+\frac{\omega}{k_R}}\right)\left(\frac{2}{1+\frac{\omega^\prime}{k_R^\prime}}\right)+
  \left(\frac{1-\frac{\omega}{k_R}}{1+\frac{\omega}{k_R}}\right)\left( \frac{1-\frac{\omega^\prime}{k_R^\prime}}{1+\frac{\omega^\prime}{k_R^\prime}}\right) \frac{k_R}{\omega}  + \frac{k_R}{\omega}    \right\}   \;.
 \eea
 Since $ \left(f_{{I}}(\omega, x), f_{{I}}(\omega^\prime, x)\right) = \delta(\w-\w')$ the above expression must be equal to 1 when $\w = \w'$.
 With this constraint we find that
\begin{align}
 N_{\omega}=\sqrt{\frac{{\frac{2 \omega^2}{k_R}+\frac{\omega^3}{k_R^2}+\omega}}{{(4 \pi \omega ) \left(\frac{\omega^2}{k_R}+k_R+2 \omega\right)}}}=\frac{1}{\sqrt{4\pi k_R}}.
\end{align}

\end{appendix}

\end{document}